\documentclass[aps,amsfonts,nofootinbib,onecolumn,showpacs]{revtex4}
\usepackage{bm}
\usepackage{graphicx}
\usepackage{amsmath}
\usepackage{epsf}
\usepackage{amssymb}
\usepackage{txfonts}
\usepackage{color}
\newcommand{\tQ}{\tilde Q}
\newcommand{\tU}{\tilde U}

\newcommand{\calb}{a}
\newcommand{\rot}{\omega}

\newcommand{\da}{d_A}

\newcommand{\tot}{{\rm t}}
\newcommand{\obs}{{\rm obs}}

\newcommand{\cmb}{T}

\newcommand{\n}{{\rm n}}
\newcommand{\vecl}{{\bf l}}
\newcommand{\vecla}{{{\bf l}_1}}
\newcommand{\veclb}{{{\bf l}_2}}
\newcommand{\veclc}{{{\bf l}_3}}
\newcommand{\vecld}{{{\bf l}_4}}

\newcommand{\intl}[1]{\int \frac{d^2 {\bf l}_#1}{(2\pi)^2}}

\newcommand{\intln}{\int \frac{d^2 {\bf l}}{(2\pi)^2}}
\newcommand{\intlnp}{\int \frac{d^2 {\bf l'}}{(2\pi)^2}}
\newcommand{\intlnpp}{\int \frac{d^2 {\bf l''}}{(2\pi)^2}}
\newcommand{\intlnppp}{\int \frac{d^2 {\bf l'''}}{(2\pi)^2}}

\newcommand{\bfl}{{\mathbf{l}}}
\newcommand{\bfd}{{\mathbf{d}}}
\newcommand{\bflp}{{\mathbf{l^{\prime}}}}
\newcommand{\bflpp}{{\mathbf{l^{\prime\prime}}}}
\newcommand{\bflppp}{{\mathbf{l^{\prime\prime\prime}}}}
\newcommand{\bfL}{{\mathbf{L}}}
\newcommand{\bfLp}{{\mathbf{L^{\prime}}}}
\newcommand{\dirac}{{\rm D}}

\newcommand{\pp}{{\phi\phi}}

\newlength{\tskip}
\newlength{\colwidth}

\newcommand{\beq}{\begin{equation}}
\newcommand{\eeq}{\end{equation}}
\newcommand{\beqa}{\begin{eqnarray}}
\newcommand{\eeqa}{\end{eqnarray}}

\newcommand{\bn}{\hat{\bf n}}
\newcommand{\bl}{\hat{\bf l}}

\newcommand{\rad}{r}    

\newcommand{\len}{\phi}

\newcommand{\est}{{{d}}_{TT}}
\newcommand{\estEB}{{{d}}_{EB}}
\newcommand{\estEE}{{{d}}_{EE}}

\newcommand{\estTX}{{{d}}_{TX}}

\newcommand{\norm}{A_{\cmb\cmb}}

\newcommand{\normEE}{A_{EE}}

\newcommand{\normTX}{A_{TX}}

\newcommand{\normXXp}{A_{XX'}}
\newcommand{\normEB}{A_{EB}}
\newcommand{\filt}{F_{TT}}
\newcommand{\ffact}{f_{TT}}
\newcommand{\ffacsys}{f_{aa}}

\newcommand{\filtEB}{F_{EB}}
\newcommand{\filtEE}{F_{EE}}

\newcommand{\filtTX}{F_{TX}}

\newcommand{\filtXXp}{F_{XX'}}
\newcommand{\noise}{N^{(0)}_{TT,TT}}
\newcommand{\noiseEB}{N^{(0)}_{EB,EB}}

\newcommand{\noiseTX}{N^{(0)}_{TX,TX}}
\newcommand{\noiseEE}{N^{(0)}_{EE,EE}}

\newcommand{\noiseP}{N^{(1)}_{TT,TT}}
\newcommand{\noiseS}{N^{(S)}_{TT,TT}}
\newcommand{\noiseSEB}{N^{(S)}_{EB,EB}}
\newcommand{\noiseSEE}{N^{(S)}_{EE,EE}}

\newcommand{\noiseSTX}{N^{(S)}_{TX,TX}}
\newcommand{\noisePEE}{N^{(1)}_{EE,EE}}
\newcommand{\noisePEB}{N^{(1)}_{EB,EB}}

\newcommand{\noisePTX}{N^{(1)}_{TX,TX}}

\newcommand{\vl}{{\mathbf{l}}}

\begin{document}

\title {Impact of Instrumental Systematic Contamination on the Lensing Mass Reconstruction using the CMB Polarization}

\author{Meng Su$^1$}\email{mengsu@cfa.harvard.edu}
\author{Amit P.S. Yadav$^1$}
\author{Matias Zaldarriaga$^{1,2}$}

\affiliation{$^1$Harvard-Smithsonian Center for Astrophysics, 60
Garden St., Cambridge, MA 02138, USA} \affiliation{$^2$Jefferson
Laboratory of Physics, Harvard University, Cambridge, MA 02138, USA
}



\begin{abstract}
In this paper, we study the effects of instrumental systematics on
the reconstruction of the deflection angle power spectrum from weak
lensing of Cosmic Microwave Background (CMB) temperature and
polarization observations. We consider 7 types of effects which are
related to known instrumental systematics: calibration, rotation,
pointing, spin-flip, monopole leakage, dipole leakage and quadrupole
leakage. These effects can be characterized by 11 distortion fields.
Each of these systematic effects can mimic the effective projected
matter power spectrum and hence contaminate the lensing
reconstruction. To demonstrate the effect of these instrumental
systematics on CMB lensing measurements, we consider two types of
experiments, one with a detector noise level for polarization of 9.6
$\mu$K-arcmin and FWHM of $8.0^\prime$, typical of upcoming ground
and balloon-based CMB experiments, and a CMBPol-like instrument with
a detector noise level for polarization of 2.0 $\mu$K-arcmin and
FWHM of $4.0^\prime$, typical of future space-based CMB experiments.
For each systematics field, we consider various choices of coherence
scale $\alpha_s$, starting from $\alpha_s=10^\prime$ to the maximum
of $\alpha_s=120^\prime$. Among all the
 11 systematic parameters, rotation $\rot$, and monopole leakage $\gamma_a \& \gamma_b$
 place the most stringent requirements, while quadrupole leakage $q$, pointing error $p_a$ and $p_b$, and
 calibration $a$ are among the least demanding.
 The requirements from lensing extraction are about 1-2 orders of
  magnitude less stringent than the
requirements to measure the primordial B-modes with inflationary
energy scale of $1.0\times10^{16}$ GeV. On the other hand the requirements
for lensing reconstruction are comparable
or even more stringent for some systematic parameters than the requirements
 to detect primordial B-modes with inflationary scale $E_i = 3.0\times10^{16}$ GeV.


\end{abstract}

\maketitle

\section{Introduction}

Observations of the temperature anisotropies of the Cosmic Microwave
Background (CMB) have been a major tool to constrain cosmological
parameters. The polarization data of the CMB can help us to extract
additional information beyond the temperature
information~\cite{Kovac,WMAP}. The next generation of CMB
observations will focus on the precise measurement of polarization
of the CMB, especially the so called B-mode polarization, which is
at least two orders of magnitude smaller than the temperature
anisotropy signal. In contrast to E-mode polarization, which can be
generated by scalar or tensor perturbations in the early universe,
the primordial B-modes are generated only by tensor
perturbations~\cite{KamKosSte97, ZalSel97}. However even in the
absence of primordial B-modes, subsequent gravitational lensing by
the large scale structure of the Universe converts E-mode
polarization to B-mode polarization~\cite{Sel96, ZalSel98, Hu00b,
others}. Although the amplitude of the primordial gravitational wave
signal is uncertain by many orders magnitude and might not be
detectable by the next generation of polarization experiments, the
lensing B-mode signal is a guaranteed prediction of the current
cosmological model. In addition, the B-mode lensing signal will help
to break degeneracies between cosmological
parameters~\cite{HHS,smith,smith2,viviana,lensinginf,secondaryBmode}.

The weak gravitational lensing of CMB
anisotropies provides a unique opportunity to map the matter
distribution of the universe.
The non-Gaussian higher order correlations in the CMB generated by
the weak lensing can be used to reconstruct the mass distribution of
the intervening large-scale structure. The principle is easy to
understand. The CMB photons are remapped by gravitational lensing
which introduces correlations between different angular moments. One
way of extracting the lensing information is to use a quadratic
combination of the CMB multipoles to define an estimator for the
projected gravitational
potential~\cite{HuOkamoto,ZalSel99,GuzSelZal00,Hirata,Hirata2,Kesden2,Kesden,Cooray}.

Although the B-mode polarization observations are currently still
noise dominated, the next generation of CMB polarization instruments
has the sensitivity to make first detections, at least of the
lensing induced B-mode signal. However, there are several challenges
for CMB lensing detection, mainly coming from astrophysical
foregrounds and instrumental systematics.
It is important
to estimate and control those spurious signals as well as possible
when analyzing upcoming CMB data. These challenges will have to be
overcome in order to probe the physics of the early universe through
$B$-mode polarization or to infer the projected large scale matter
distribution from E/B polarization.

Lensing studies can be considered as secondary science for an
experiment devoted to B-mode detection. Impact of instrumental
systematics on the projected matter power spectrum
is helpful to both instrument design and future data forecast. On
the other hand, lensing induced $B$-mode polarization is a
cosmological contaminant for the detection of primordial $B$-modes,
a systemic study of instrumental systematics for lensing
reconstruction may well be required of such an experiment to delense
the observed CMB
fields~\cite{Hu01c,CMBpol,EBEX,Spider,Task,QUIJOTE,Capmap,EPIC,experiments}.

In the literature, instrumental systematics have been discussed
extensively
~\cite{HHZ,DCJ,Shimon,Shimon2,Bunn,Beammismatch,modulation,beamasym}.
The main goal of this paper is to illustrates the effects of
instrumental systematics and systematically study the impact on the
mass reconstruction process for upcoming CMB experiments~\cite{EBEX,
CMBpol,Spider,QUIJOTE, Capmap, Task, EPIC}. To calculate the effects
of instrumental systematics on the projected matter power spectrum,
we make use of the quadratic estimator to reconstruct the projected
gravitational potential \cite{HuOkamoto}, and classify and
parameterize the systematics following~\cite{HHZ}. We divide
polarization contaminations into two categories: those which are
associated with a transfer between the polarization state of the
incoming radiation (from detection system), including calibration
and rotation, spin-flip coupling and monopole leakage errors, and
those which are associated with CMB anisotropy induced by the finite
resolution or beam of the telescope. For the polarization transfer
systematics, we only concern ourselves with polarization transfer in
a single, perfectly known, direction on the sky. However, in
reality, every experiment necessarily has finite resolution and this
therefore is an additional class of contamination associated with
the resolution or beam of the experiment. We refer the reader
to~\cite{HHZ} for a detailed discussion of the parametrization of
the systematic errors we use in this paper. We calculate the
unlensed and lensed CMB power spectrum using CMBFAST~\cite{CMBFAST}.
In the calculation,  we have assumed a flat $\Lambda CDM$ cosmology
with following cosmological parameter values: $\Omega_b=0.045,
\Omega_c=0.23, H_0=70.5, n_s=0.96, n_t=0.0, \tau=0.08$.

This paper is organized as follows: In Sec.~\ref{lensing_formalism}
we review the basic lensing formalism, define our notation, and
introduce the quadratic estimator of the lensing potential
reconstruction which we later use to explore the effects of
instrumental systematics. In Sec.~\ref{systematics}, we first
consider a relatively simple case of the calibration systematics on
the temperature estimator of lensing reconstruction. Then we
consider EB estimator as an example, studying 7 types of
instrumental systematic effects on lensing reconstruction, and
compare to the systematic contamination of the B-mode power spectrum
detection. We consider two instrumental configurations, one with
noise sensitivity
 for polarization of 9.6 $\mu$K-arcmin and FWHM of $8.0^\prime$ (Exp1 from here on),
and another CMBPol like instrument with noise sensitivity for
polarization of 2.0 $\mu$K-arcmin and FWHM of $4.0^\prime$
(reference experiment from here on). In Sec.~\ref{results}
and~\ref{conclusion}, we describe our results and conclude with a
discussion of the implications for experiments dedicated to measure
primordial B-modes or the secondary lensing signal. We leave
discussions of the other three lensing potential estimators ($ EE,
TE, TB$) to Appendix.

\section{Lensing Formalism}

\label{lensing_formalism}

Gravitational lensing deflects the path of CMB photons from the last
scattering surface resulting in a remapping of the CMB
temperature/polarization pattern on the sky. In this section, we
review the basic lensing calculation for both temperature and
polarization fields as the starting point of our discussion. We
formulate CMB lensing using the flat-sky
approximation~\cite{Cooray}. The flat-sky approach simplifies the
derivation by replacing summations over Wigner symbols of spherical
harmonic moments by integrals involving mode coupling
angles~\cite{Hu00b}. More details of CMB lensing can be found in the
nice review paper~\cite{review}.

At a certain position $\bn$ on the sky, the observed CMB field
$\tilde X(\bn)$ is lensed from another direction in the primordial
CMB sky $X(\bn + \bfd(\bn))$ at z=1090. The remapping process can be
described as

\begin{eqnarray}
\tilde\cmb(\bn) & = &  \cmb(\bn + \bfd(\bn)) \,,\\
\, [\tilde Q\pm i\tilde U](\bn) & = &  [Q\pm i  U](\bn +\bfd(\bn))\,, \nonumber
\label{E:RealTay}
\end{eqnarray}
where $\tilde\cmb(\bn)$ ($\cmb(\bn)$) represents the lensed
(unlensed) temperature fluctuation field, $\tilde Q(\bn)$ ($Q(\bn)$)
and $\tilde U(\bn)$ ($U(\bn)$)
 are lensed (unlensed) polarization Stokes parameters, and $\bfd(\bn)$ is the deflection angle which is related to
$\len(\bn)$,
 the lensing gravitational potential, by $d(\bn)$ = $\nabla\len(\bn)$. Here and throughout this paper, we use boldface quantities to identify vectors,
 and $\tilde X$ ($X$) stands for lensed (unlensed) temperature and polarization fields. The
 lensing potential $\len(\bn)$ is given by

\begin{eqnarray}
\phi(\bn)&=&- 2 \int_0^{\rad_0} d\rad
\frac{\da(\rad_0-\rad)}{\da(\rad)\da(\rad_0)} \Phi (\rad,\rad\bn )
\, , \label{eqn:lenspotential}
\end{eqnarray}
where $d_A$ is the comoving distance along the line of sight; $r_0$ is the comoving distance to the surface of last scattering, and $\Phi$ is
gravitational potential. The lensing remapping process
conserves the surface brightness distribution of the CMB, thus does not
change the one-point statistics.

The observed temperature and polarization fluctuations also include
secondary effects, such as Sunyaev-Zel'dovich (SZ) effect~\cite{SZ}
and Integrated Sachs Wolfe (ISW) effect~\cite{ISW}, which come from
the first order density or potential fluctuation and thus also
correlate with the lensing deflection angle. We denote these
physical contamination to lensing reconstruction by $X^{sec}(\bn)$.
We denote the noise component by $X^{n}(\bn)$. The total observed
CMB anisotropy therefore includes the lensed primary signal, any
secondary effects, and noise, i.e. $X^{t}(\bn) = \tilde X(\bn) +
X^{sec}(\bn) + X^{n}(\bn)$. In the next section, we will introduce
another contribution to $X^{t}(\bn)$, which comes from instrumental
systematics $X^{sys}(\bn)$. We define the observed CMB field
$X^{obs}(\bn) = X^{t}(\bn) + X^{sys}(\bn)$. Here we write the
secondary contribution as an independent component from the lensed
CMB. However, in reality it is hard to separate $X^{sec}(\bn)$ from
$X^{t}(\bn)$ because secondaries are also lensed by gravitational
potentials with deflection angles depending on their redshifts. In
this paper, we simply drop the contribution effects $X^{sec}(\bn)$,
as this topic is beyond the focus of this paper. We refer the
readers to \cite{Cooray} for a treatment of the secondary anisotropy
as a physical contamination to the lensing potential reconstruction
analysis\footnote{ As a note, thermal SZ effect can in principal be
separated from the primary fluctuations by its spectral dependence.
For the kinetic SZ effect, it was claimed that by using a specially
designed estimator~\cite{ksz}, it is possible to separate it out
from real lensing signal. However, some important secondary
contributions such as the ISW effect cannot be separated easily and
will lead to additional noise contributions due to correlations with
the lensing potentials~\cite{Cooray}.}.

 It is convenient to work in Fourier space. If one considers a
small enough patch of sky, spherical harmonic modes can be replaced
by Fourier modes.
Generalization from the flat-sky to the full sky is straightforward. The Fourier transform of the Taylor expended lensed CMB temperature
and polarization field is
\begin{eqnarray}
\tilde \cmb(\bfl)
&=&\int d \bn\, \tilde \cmb(\bn) e^{-i \bfl \cdot \bn}
= \cmb(\bfl) - \intlnp \cmb(\bflp) L(\bfl,\bflp), \\
 \left[ \tilde E\pm i \tilde B \right] (\bfl) &=&
        \int  d \bn\, [\tilde Q(\bn)\pm i \tilde U(\bn)] e^{\mp 2i\varphi_{\bf l}} e^{-i \bl \cdot \bn}=[E(\bfl)\pm i B(\bfl)]- \intlnp [E(\bflp)\pm i B(\bflp)] L_p(\bfl,\bflp) \,,
        \nonumber\\
\phi(\bfl) &=&\int d \bn\,  \phi(\bn) e^{-i \bfl \cdot \bn}\,,
\label{E:thetal}
\end{eqnarray}
where
\begin{eqnarray}
\label{E:lfactor} L(\bfl,\bflp) &\equiv& \len(\bfl-\bflp) \, \left[
(\bfl-\bflp) \cdot \bflp \right] +\frac{1}{2} \intlnpp \len(\bflpp)
\times\len(\bfl - \bflp - \bflpp) \, (\bflpp \cdot
\bflp)
                \left[ (\bflpp + \bflp - \bfl)\cdot
                             \bflp \right] + \ldots \,,\\
L_P(\bfl,\bflp) &\equiv&
e^{\pm2i(\varphi_{\vl'}-\varphi_\vl)}\len(\bfl-\bflp) \, \left[
(\bfl-\bflp) \cdot \bflp \right] + \frac{1}{2} \intlnpp
e^{\pm2i(\varphi_{\vl'}-\varphi_\vl)} \len(\bflpp)\times\len(\bfl - \bflp
- \bflpp) \, (\bflpp \cdot
\bflp)\left[ (\bflpp + \bflp - \bfl)\cdot \bflp \right] + \ldots \,.\nonumber
\end{eqnarray}

We can immediately see that lensing induces remapping of CMB fields
by lensing potential gradients, hence in Fourier space lensing acts
as a convolution which couples different harmonic modes. Fourier
moments, power spectrum, bispectrum, trispectrum and so on of the
CMB fields and the lensing potential can be defined in the usual
manner:
\begin{eqnarray} \label{E:correlationdef}
\left< X^i(\bfl_1) X^{\prime j}(\bfl_2)\right> &\equiv&
        (2\pi)^2 \delta_\dirac(\vecl_1+\vecl_2)  C^{ij}_{X^{i}X^{\prime j}}(l_1)\,,\nonumber\\
\left< X^i(\bfl_1) X^{\prime j}(\bfl_2)X^{\prime\prime k}(\bfl_3)\right>_c &\equiv& (2\pi)^2 \delta_\dirac(\vecl_1+\vecl_2+\vecl_3)
        B^{ijk}_{\rm{XX^{\prime}X^{\prime\prime}}}(\bfl_1,\bfl_2,\bfl_3)\,,\nonumber\\
\left< X^i(\bfl_1) X^{\prime j}(\bfl_2)X^{\prime\prime k}(\bfl_3)X^{\prime\prime\prime m}(\bfl_4)\right>_c&\equiv& (2\pi)^2 \delta_\dirac(\vecl_1+\vecl_2+\vecl_3+\vecl_4)
        T^{ijkm}_{XX^{\prime}X^{\prime\prime}X^{\prime\prime\prime}}(\bfl_1,\bfl_2,\bfl_3,\bfl_4)\,,\nonumber\\
 && \quad       \ldots \end{eqnarray}
where the angle brackets represent ensemble averages over
realizations of the primordial CMB fields, the large-scale structure
between observers and the last scattering surface, and the
experimental noise. The connected part of the n-point function is
denoted by the subscript $c$. The fields $X$, $X^{\prime}$,
$X^{\prime\prime}$, $X^{\prime\prime\prime}$ are among {$\{
\cmb(\vecl)$, $E(\vecl)$, $B(\vecl)$, $\phi(\vecl)\}$}. The
superscripts $i,j,k,m$ represent the unlensed field $X$, the lensed
field $\tilde X$, the instrumental noise $X^{n}$, the instrumental
systematics $X^{sys}$, CMB secondary contribution $X^{sec}$, the
total signal $X^{t}$, or the observed field $X^{obs}$. We note that
the bispectrum and higher order odd-correlations vanish if one
ignores the secondary effects. This is because odd moments contain
sample averages over the odd primordial CMB fields which we assume
to be Gaussian.

We make the assumption that fluctuations in the large-scale
structure between the observer and the last scattering surface are
Gaussian and hence can be fully described by a power spectrum. We
use the lensing potential power spectrum calculated from CAMBFAST.
The instrumental noise $X^{n}$ is also assumed to be Gaussian. We
note that primordial non-Gaussianity can possibly contribute percent
level uncertainty to our analysis which is considered in
\cite{nonGaussianity}.

We are now in the position to calculate any order (cross)
correlation functions of CMB fields and the lensing potential in
Fourier space. We will assume uniform Gaussian noise with the power
spectrum $C_l^{{XX}\n}$ given by
\begin{eqnarray} \label{E:noise}
C_l^{XX\n} = w_X^{-1} e^{l^2 \sigma_{b}^2} \,.
\end{eqnarray}
$w^{-1}_X$ is the detector noise variance per steradian area for
temperature (X = T) or polarization (X = E or B), and $\sigma_b =
\theta_{fwhm}/\sqrt{8 \ln 2}$ is the effective beamwidth of the
instrument calculated from its full-width half-maximum resolution
$\theta_{fwhm}$. We will assume fully polarized detector, for which
$2w_T=w_E=w_B$.

\begin{figure}[]
\centering
\includegraphics[scale=0.90] {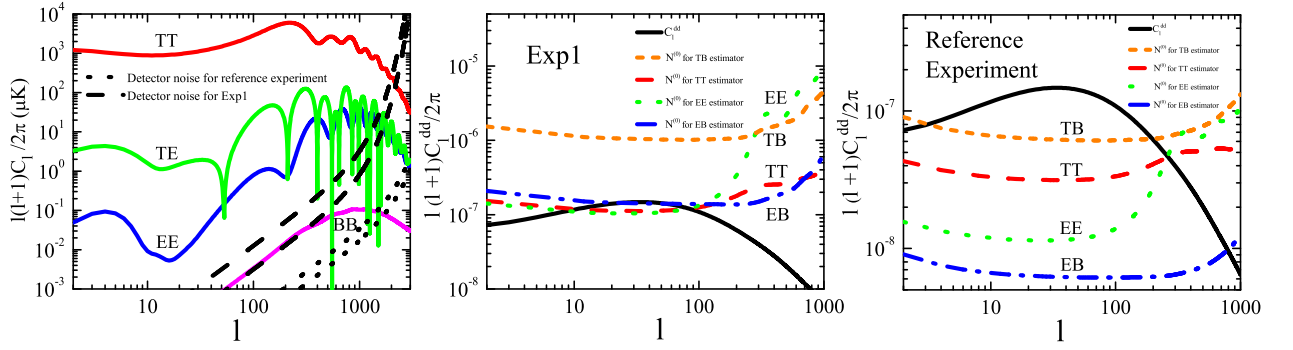}
\caption{{\it Left panel}: CMB power spectrum for the TT, TE, EE and
lensed BB power spectrum. The lower (upper) dashed black line shows
the temperature (polarization) noise for Exp1. The lower (upper)
dotted black line shows the temperature (polarization) noise for
CMBPol-like reference experiment. {\it Center panel}: Gaussian noise
for different quadratic estimators for the Exp1. {\it Right panel}:
Same as central panel but for CMBPol-like reference experiment. Note
that the EB estimator has the lowest Gaussian noise, and may thus be
considered as the best estimator among all the possible quadratic
estimators. } \label{fig:gaussian}
\end{figure}

Quadratic combinations of CMB fields can be used as estimators of the
 lensing potential field and hence the intervening projected mass between us and the last scattering surface.
 Furthermore, a CMB-field-squared map appropriately filtered in Fourier space can serve as an optimal estimator
  for the deflection field. Optimal filters for quadratic estimators have been designed~\cite{HuOkamoto}
\begin{equation} \label{E:estXX}
d_{XX^\prime}(\bfL) \equiv \frac{A_{XX'}(L)}{L} \intl{1}
X^\tot(\vecla) X'^\tot(\veclb) \filtXXp(\vecla, \veclb) \,,
\end{equation}
where $X$ and $X^\prime$ can be T, E, and B. The normalization
$A_{XX^\prime}$ is chosen such that
 $\langle d_{XX^\prime}(\bfL)\rangle_{\rm{CMB}} = d(\bfL)\equiv L\phi $

\begin{equation} \label{E:normXX}
\normXXp(L) \equiv L^2 \left[ \intl{1} f_{XX^\prime}(\vecla, \veclb)
\filtXXp(\vecla, \veclb) \right]^{-1} \, ,
\end{equation}
 where

\begin{eqnarray}
F_{XX^\prime}(\vecla, \veclb) &=& \frac{ C_{l_1}^{X^\prime X^\prime \tot} C_{l_2}^{XX \tot} f_{XX^\prime}(\vecla, \veclb)-C_{l_1}^{XX^\prime \tot} C_{l_2}^{XX^\prime \tot}
f_{XX^\prime}(\veclb, \vecla)} {C_{l_1}^{XX\tot} C_{l_2}^{X^\prime X^\prime \tot} C_{l_1}^{X^\prime X^\prime \tot} C_{l_2}^{XX\tot} - (C_{l_1}^{XX^\prime \tot} C_{l_2}^{XX^\prime \tot})^2}
\, ,
\label{eq:F}
\end{eqnarray}
where for $XX' = \cmb \cmb, EE, BB$, and $\cmb E$,
\begin{eqnarray}
f_{X X'}(\bfl_1, \bfl_2) = C_{l_1}^{X X'} \, ^{1}W_{X
X'}(\vl_1,\vl_2) + C_{l_2}^{X X'} \, ^{2}W_{X X'}(\vl_1,\vl_2),
\label{eq:f1}
\end{eqnarray}
and for $X = \{\cmb, E\}$, $X' = B$,
\begin{eqnarray}
f_{X X^\prime}(\bfl_1, \bfl_2) = C_{l_1}^{X E} \, ^{1}W_{X
X^\prime}(\vl_1,\vl_2) + C_{l_2}^{X^\prime X^\prime} \, ^{2}W_{X
X^\prime}(\vl_1,\vl_2). \label{eq:f2}
\end{eqnarray}
The window functions $W_{X X^\prime}$ are given in
Table~\ref{tab_wxx}. In Fig.~\ref{fig:gaussian}, we show the input
power spectrum of the lensed CMB fields, the reconstructed
deflection field, and the corresponding Gaussian noise as obtained
by using the estimators defined in Eq. (\ref{E:estXX}). Note that
the average $\langle \quad \rangle_{\rm{CMB}}$ denotes an ensemble
average restricted only to different Gaussian realizations of the
primordial CMB and instrument noise but assuming a fixed realization
of the large-scale structure. The unmarked average, $\langle \quad
\rangle$, means the average over the primordial CMB field \emph{and}
the large-scale structure realizations as defined in
Eq.~(\ref{E:correlationdef}). For the purposes of estimating the
large-scale structure in the real observable universe, it is
essential to ensure that the estimators after appropriate averaging
over realizations are truly unbiased for a typical realization of
the primordial Gaussian CMB field. As we will see in the next
section that in the presence of non-zero systematic contamination,
the estimators are biased.

\begin{table}
\begin{tabular}{|c|c|c|}
\hline\hline
$X X^\prime$&$^1W_{X X^\prime}(\vl_1,\vl_2)$&$^2W_{X X^\prime}(\vl_1,\vl_2)$\\
\hline
$\cmb \cmb$& $(\bfL \cdot \bfl_1)$  & $(\bfL \cdot \bfl_2)$ \\
$\cmb E$& $\cos 2 (\varphi_{\bfl_1}
-\varphi_{\bfl_2}) (\bfL \cdot \bfl_1)$ & $(\bfL \cdot \bfl_2)$\\
$\cmb B$& $\sin 2 (\varphi_{\bfl_1}
-\varphi_{\bfl_2}) (\bfL \cdot \bfl_1)$ & 0\\
$EE$& $\cos 2 (\varphi_{\bfl_1} -\varphi_{\bfl_2})(\bfL \cdot
\bfl_1)$ & $\cos 2
(\varphi_{\bfl_1} -\varphi_{\bfl_2}) (\bfL \cdot \bfl_2)$ \\
$EB$& $\sin 2
(\varphi_{\bfl_1} -\varphi_{\bfl_2})(\bfL \cdot \bfl_1) $ & $\sin 2
(\varphi_{\bfl_1} -\varphi_{\bfl_2}) (\bfL \cdot \bfl_2)$\\
$BB$&  $\cos 2
(\varphi_{\bfl_1} -\varphi_{\bfl_2}) (\bfL \cdot \bfl_1)$& $\cos 2
(\varphi_{\bfl_1} -\varphi_{\bfl_2}) (\bfL \cdot \bfl_2)$\\
\hline
\end{tabular}
\caption{Window functions which appear in Eq.~(\ref{eq:f1}) and
Eq.~(\ref{eq:f2}); here $\bfL = \bfl_1 + \bfl_2$ } \label{tab_wxx}
\end{table}

\section{Instrumental Systematics Effect on the Lensing Potential Power Spectrum}
\label{systematics}

In this section, we use the quadratic estimators to derive the
effects of instrumental systematics on the reconstructed lensing
potential power spectrum. We show that instrumental systematics can
introduce non-Gaussian correlations of CMB fields. The instrumental
systematics-induced CMB trispectrum gives an extra contamination to
the reconstructed deflection angle power spectrum. In
subsection~\ref{temsystematics}, we consider the simple case of the
TT estimator to explain how systematics contaminate the lensing
reconstruction process. In subsection~\ref{polasystematics}, we take
the EB estimator as an example to show how instrumental systematics
in CMB polarization measurements affect the lensing reconstruction.
In order to compare the experimental requirements for primordial
B-mode detection and lensing potential reconstruction, we also
calculate the effects of instrumental systematics on B-mode
detection for a given inflationary energy scale.

\subsection{A simple example of temperature systematics}
\label{temsystematics}

We introduce the calibration parameter (gain fluctuation of
receivers) $a(\bn)$ for temperature measurement, defined as:
\begin{equation}
\tilde\cmb^{obs}(\bn) =  [1+a(\bn)]\tilde\cmb^{t}(\bn)\,.
\label{E:expandsys}
\end{equation}

If we assume that there is no correlation between the lensing
potential $\phi$ and the instrumental systematic $a(\bn)$, the power
spectrum of the lensed CMB temperature with systematics correction
term can be found as (more details are given in
Appendix~\ref{appentemsys}):

\begin{eqnarray}
\tilde C_l^{\cmb\cmb} &=& \left[ 1 - \intl{1} C^{\phi\phi}_{l_1}
\left(\vecl_1 \cdot \vecl\right)^2 \right]   \,
                            C_l^{\cmb\cmb}
        + \intl{1} C_{| \vecl - \vecl_1|}^{\cmb\cmb} C^{\phi\phi}_{l_1}
                [(\vecl - \vecl_1)\cdot \vecl_1]^2  \, + \intl{1} C_{| \vecl - \vecl_1|}^{aa} C^{\cmb\cmb}_{l_1}\,.
\label{E:lenpower}
\end{eqnarray}
This result is given to linear order in the lensing-potential power
spectrum $C^{\phi\phi}_l$ and the gain fluctuation systematics power
spectrum $C^{aa}_l$. The last term represents the bias introduced by
the calibration systematics. In the absence of systematic effects,
it is easy to prove that the deflection angle estimator is $\langle
\est(\bfL) \rangle_{\rm{CMB}} = L \len(\bfL)\equiv \est(\bfL)$, as
desired. But once the contribution from the calibration parameter
$a(\bn)$ is considered, one finds:

\begin{equation}
\langle \est(\bfL) \rangle_{\rm{CMB}}= \est(\bfL)+ \frac{
\norm(L)}{L} \intl{1}
\filt(\vecla,\veclb)a(\bfL)\,(C_{l_1}^{\cmb\cmb}\,+C_{l_2}^{\cmb\cmb})\,,
\end{equation}
i.e. in the presence of $a(\bfL) $ the estimator $\est(\bfL)$ is a
biased estimator for the deflection field in Fourier space.
Consequently, the deflection angle power spectrum $\langle
\est(\bfL) \est(\bfL)\rangle$ would be biased due to systematic
contamination, and is given by (see Appendix~\ref{appentemsys} for
details)

\begin{eqnarray}
\Big\langle\Big\langle \langle \est(\bfL) \cdot
\est(\bfLp)\rangle_{\rm{CMB}}\Big\rangle_{\rm{LSS}}\Big\rangle_{\rm{SYS}}
&=& \frac{\norm(L)}{L}
\frac{\norm(L^{\prime})}{L^{\prime}} \nonumber \\
&& \quad \times \intl{1} \intl{1'}(2 \pi)^2 \filt(\vecla, \veclb) \filt(\vecla', \veclb')\nonumber \\
&& \quad \Bigg\{C_{L}^\pp \ffact(\vecla,\veclb)
\ffact(\vecla',\veclb')\delta_\dirac(\bfL + \bfLp) \nonumber \\
&& \quad + (2 \pi)^2 C_{l_1}^{\cmb\cmb\tot} C_{l_2}^{\cmb\cmb\tot}
\Big[ \delta_\dirac(\vecla' + \vecla) \delta_\dirac(\veclb' +
\veclb)
 + \delta_\dirac(\veclb' + \vecla) \delta_\dirac(\vecla' +
\veclb) \Big] \nonumber \\
&& \quad + \Big[C_{|\vecla+\vecla'|}^\pp \ffact(\vecla,\vecla')
\ffact(\veclb,\veclb') + C_{|\vecla+\veclb'|}^\pp
\ffact(\vecla,\veclb')
\ffact(\veclb,\vecla')\Big]\delta_\dirac(\bfL + \bfLp) \nonumber \\
&& \quad + \Big[ C_{L}^{aa} \ffacsys(\vecla,\veclb)
\ffacsys(\vecla',\veclb') + C_{|\vecla + \vecla'|}^{aa}
\ffacsys(\vecla,\vecla') \ffacsys(\veclb,\veclb') \nonumber \\
&& \quad + C_{|\vecla + \veclb'|}^{aa}\ffacsys(\vecla,\veclb')
\ffacsys(\veclb,\vecla') \Big] \delta_\dirac(\bfL + \bfLp) \Bigg\} \nonumber \\
&=& (2 \pi)^2 \delta_\dirac(\bfL + \bfLp) \Bigg[ C_{L}^{dd}+
\noise(L) + \noiseP(L) + \noiseS(L) + ...\Bigg], \label{E:temsys}
\end{eqnarray}

where we define $\ffacsys(\vecla,\veclb) = C_{l_1}^{\cmb\cmb} +
C_{l_2}^{\cmb\cmb}$. In the last line, the first term in the square
bracket is the deflection angle power spectrum $C_{L}^{dd}$. The
second term is the so called Gaussian noise $\noise(L)$ which gives
the dominant noise contribution to the variance of the deflection
power spectrum. The third term $\noiseP(L)$ is the the leading order
non-Gaussian noise which is first order in $C_{L}^\pp$ and gives
correction to the dominant Gaussian noise $\noise(L)$. The forth
term, $\noiseS(L)$, is the leading order instrumental systematic
contribution to the variance and is first order in $C_{L}^{SS}$. The
Gaussian noise $\noise(L)$ and the first order non-Gaussian noise
$\noiseP(L)$ have been previously calculated in~\cite{HuOkamoto}
and~\cite{Kesden}, respectively. The systematic noise term
$\noiseS(L)$ is a new contribution to lensing power spectrum. In
principle, one should include noise terms which are higher order in
$C_{L}^\pp$ and $C_{L}^{SS}$, however since both of them are small,
we truncate at the first order and expect that higher order
contributions are much smaller.

The quadratic estimator given in Eq.~(\ref{E:estXX}) is optimized in
the presence of $\noise(L)$, and assuming no contribution from the
first order non-Gaussian noise $\noiseP(L)$ and instrumental
systematic noise $\noiseS(L)$. Hence the estimator is
optimal\footnote{If the systematic contributions are comparable to
the Gaussian noise contribution then we need to design new optimal
lensing reconstruction estimators to take into account the
instrumental systematics effect.} as long as $\noiseP(L) \ll
\noise(L)$, and $\noiseS(L) \ll \noise(L)$. It has been shown that
the non-Gaussian noise $\noiseP(L)$ is about one order of magnitude
smaller than the Gaussian noise contribution~\cite{Kesden}. In
\ref{polasystematics}, we calculate the systematic noise term
$N^{S}(L)$ contribution to the lensing potential reconstruction for
the EB estimator.

\subsection{General analysis on polarization systematics}
\label{polasystematics}

We parametrize the fields of instrumental systematics for CMB polarization measurements following~\cite{HHZ}. The polarization contaminations
fall into two categories, one associated with the detector system which distorts the
polarization state of the incoming polarized signal (Type I hereafter), and another associated with distortion of the CMB signal due to the beam anisotropy (Type II hereafter). This parametrization can be generalized to different polarimeters.
The instrumental response to incoming CMB radiation is usually described by the Jones transfer matrix. Bias induced in the matrix determination will mix the Stokes parameters determined from it. To first order, the effect of Type I systematics on the Stokes parameters can be written as~\cite{HHZ}

\begin{equation}
\delta [Q \pm i U](\bn) =
            [\calb \pm i 2 \rot](\bn)  [Q \pm i U](\bn) + [f_1 \pm i f_2](\bn)   [Q \mp i U](\bn) + [\gamma_1 \pm i \gamma_2](\bn) \cmb(\bn).\,
\label{eq:pointsys}
\end{equation}
$a$ is a scalar field which describes the miscalibration of the polarization measurements (recall that in last subsection,
we used $a$ to denote the miscalibration of temperature measurements), $\rot$ is also a scalar field that describes the rotation
angle of the instrument, $(f_1\pm if_2)$ are spin $\pm 4$ fields that describe the coupling between two spin states (spin-flip),
 and $(\gamma_1\pm i \gamma_2)$ are spin $\pm2$ fields that describe
monopole leakage from the temperature to polarization.

Similar to the Type I systematics, the effect of Type II systematics on
the Stokes parameters can be written as~\cite{HHZ}
\begin{equation}
\label{eq:localmodel}
\delta[Q \pm i U](\bn;\sigma) = \sigma {\bf p}(\bn) \cdot \nabla [Q \pm i U](\bn;\sigma) + \sigma [d_1 \pm i d_2](\bn) [\partial_1 \pm i\partial_2] \cmb(\bn;\sigma)
 + \sigma^2 q(\bn) [\partial_1 \pm i \partial_2]^2 \cmb(\bn;\sigma)\,
\end{equation}
the systematic fields are smoothed over the average beam $\sigma$ of the
experiment.  Therefore the type II systematic fields are sensitive to the imperfection of the beam on the scale $\sigma$.
$(p_1\pm ip_2)$ are spin $\pm 1$ fields that describe pointing errors,
$(d_1\pm id_2)$ are also spin $\pm1$ fields that describe dipole leakage
from temperature to polarization, and $q$ is a scalar field that describes quadrupole leakage~\cite{HHZ}.

As a simple model, we will assume that the contamination
fields, as defined in (\ref{eq:pointsys}) and (\ref{eq:localmodel}),
are statistically isotropic and Gaussian (although some of the
systematics fields need not be so), thus their statistical
properties can be fully described by their power spectra,
\begin{equation}
\left< S (\bfl) S(\bfl') \right> = (2\pi)^2 \delta(\bfl+\bfl') C_l^{S
S}\,,
\end{equation}
where $S$ stands for any of the 11 systematic fields. The systematic
fields can be modeled with the power spectra of the form
\begin{equation}
C_l^{SS} = C_{0} \exp(-l(l+1)\alpha_{S}^2), \label{eq:coh}
\end{equation}
i.e. white noise above
certain coherence scale $\alpha_{S}$, which is a key quantity to
affect the level of contamination of each systematics effects. The
normalization factor $C_{0}$ can be determined by

\begin{equation}
C_{0}^2 = A_S^2 \Bigg[\int {d^2 l \over (2\pi)^2}
\exp(-l(l+1)\alpha_{S}^2)\Bigg]^{-1}\,, \label{eq:rms}
\end{equation}
where $A_S$ characterizes the $\it rms$ of the contamination field
$S$.

The instrumental systematics induce distortions on
 the CMB fileds. The contaminations to the $BB$ and $EE$ power
spectra due to different measurement systematics take the form
\begin{equation}
\delta C_l^{BB} = \sum_{SS'} \int {d^2 \vl' \over (2\pi)^2}
C_{|\vl-\vl'|}^{SS'} C_{|\vl'|}^{EE}(\sigma)
 [W^{S}_{B}(\vl, \vl')]^2\,+ \sum_{SS'} \int {d^2 \vl' \over (2\pi)^2}
C_{|\vl-\vl'|}^{SS'} C_{|\vl'|}^{\cmb\cmb}(\sigma)
 [W^{S}_{B}(\vl, \vl')]^2\,,
 \label{eq:Bmodesys}
\end{equation}

\begin{equation}
\delta C_l^{EE} = \sum_{SS'} \int {d^2 \vl' \over (2\pi)^2}
C_{|\vl-\vl'|}^{SS'} C_{|\vl'|}^{EE}(\sigma)
 [W^S_{E}(\vl, \vl')]^2\,+ \sum_{SS'} \int {d^2 \vl' \over (2\pi)^2}
C_{|\vl-\vl'|}^{SS'} C_{|\vl'|}^{\cmb\cmb}(\sigma)
 [W^S_{E}(\vl, \vl')]^2\,.
 \label{eq:Emodesys}
\end{equation}

The explicit forms of $W^S_{B}(\vl_1,\vl_2)$ and
$W^S_{E}(\vl_1,\vl_2)$ are given in Table~\ref{table:geometricd},
which are the window functions of each systematic S for B-mode and
E-mode harmonics, respectively. The summations of the first term on
the RHS of Eq. (\ref{eq:Bmodesys}) and Eq. (\ref{eq:Emodesys}) run
over calibration $a$, rotation $\rot$, spin flip $f_a$ and $f_b$,
and pointing error $\gamma_a$ and $\gamma_b$. The summations of the
second term run over the rest of the systematics parameters which
describe the temperature leakage given in
Table~\ref{table:geometricd}. $C_{l}^{EE}(\sigma)$ and
$C_{l}^{\cmb\cmb}(\sigma)$ are the beam smoothed temperature and
E-mode polarization power spectra.

\begin{eqnarray}
C_{l}^{EE}(\sigma) = C_l^{EE} \exp(-l(l+1)\sigma), \qquad
C_{l}^{\cmb\cmb}(\sigma) = C_l^{\cmb\cmb} \exp(-l(l+1)\sigma).
\end{eqnarray}



\begin{table}
\begin{tabular}{||c||c|c||}
\hline\hline
\bf {Type of S}&$W^{S}_{B}(\bfl_1,\bfl_2)$&$W^{S}_{E}(\bfl_1,\bfl_2)$ \\
\hline
Calibration $a$ & $\sin[2 (\varphi_{l_2} - \phi_L)]$ & $\cos[2 (\varphi_{l_2} - \varphi_L)]$\\
Rotation $\rot$ & $2 \cos[2 (\varphi_{l_2} - \varphi_L)]$ & $-2 \sin[2 (\varphi_{l_2} - \varphi_L)]$ \\
Pointing $p_a$   & $\sigma (\vl_2 \times \hat \vl_1)\cdot \hat{\bf z} \sin[ 2(\varphi_{l_2}- \varphi_L) ]$ & $\sigma (\vl_2  \cdot \hat \vl_1) \sin[ 2(\varphi_{l_2} - \varphi_l)]$ \\
Pointing $p_b$   & $\sigma (\vl_2  \cdot \hat \vl_1) \sin[ 2(\varphi_{l_2} - \varphi_l)]$ & $-\sigma (\vl_2 \times \hat \vl_1)\cdot \hat{\bf z} \sin[ 2(\varphi_{l_2}- \varphi_L) ]$\\
Flip $f_a$   & $\sin[2 (2 \varphi_{l_1} - \varphi_{l_2} -\varphi_{L}) ]$ & $\cos[2 (2 \varphi_{l_1} - \varphi_{l_2} -\varphi_{L}) ]$\\
Flip $f_b$   & $\cos[2 (2 \varphi_{l_1} - \varphi_{l_2} -\varphi_{L}) ]$ & $-\sin[2 (2 \varphi_{l_1}-\varphi_{l_2}-\varphi_{L})]$ \\
Monopole $\gamma_a$ & $\sin[2 (\varphi_{l_1}- \varphi_l)]$&$\cos[2 (\phi_{l_1}- \varphi_l)]$ \\
Monopole $\gamma_b$ & $\cos[2 (\varphi_{l_1}- \varphi_l)]$& $-\sin[2 (\varphi_{l_1}- \phi_l)$ \\
Dipole $d_a$      & $- (l_2 \sigma) \cos[ \varphi_{l_1} + \varphi_{l_2} - 2 \varphi_l]$& $(l_2 \sigma) \sin[ \varphi_{l_1} + \phi_{l_2} - 2 \varphi_l]$ \\
Dipole $d_b$      &   $(l_2 \sigma) \sin[ \varphi_{l_1} + \phi_{l_2} - 2 \varphi_l]$& $ (l_2 \sigma) \cos[ \varphi_{l_1} + \varphi_{l_2} - 2 \varphi_l]$ \\
Quadrupole $q$       &$ - (l_2 \sigma)^2 \sin[ 2 (\varphi_{l_2} - \varphi_l)]$ &$ - (l_2 \sigma)^2 \cos[ 2 (\varphi_{l_2} - \varphi_l)]$ \\
\hline
\end{tabular}
\caption{Window functions for all the 11 systematic parameters.
First column indicates the type of systematic parameters in
consideration. Second and third columns show window functions for
systematics induced B-mode $W^{S}_{B}(\bfl_1,\bfl_2)$, and for
E-mode $W^{S}_{E}(\bfl_1,\bfl_2)$ respectively. These window
functions are needed to calculate systematic contamination on
primordial gravitational wave detection or the deflection angle
power spectrum reconstruction. We note that ${\vl_1} = l_1 \bl_1,\,
\vl_2 = \bfL - \vl_1$, and ${\vl_2} = l_2 \bl_2$.}
\label{table:geometricd}
\end{table}

We use Eq. (\ref{eq:Bmodesys}) and Eq. (\ref{eq:Emodesys}) to calculate
the systematic requirements for B-mode detection in order to compare
the requirements for lensing reconstruction. We show the results in
Tables~\ref{table:syst} and~\ref{table:Ecrit}.

Now we move on to calculate the systematic contamination on the
lensing power spectrum. The polarization fields are essentially
uncorrelated with the lensing potential, so if we do not consider
secondary effects, the $n$-point functions with $n$ odd are zero.
The next non-zero order is the trispectrum. The calculation for the
connected part of the trispectrum of polarization in the presence of
instrumental systematics is similar to the connected temperature
trispectrum presented in the last section. Here we give the results
for the trispectrum related to lensing reconstruction using EB
estimator, and refer the readers to the
Appendix~\ref{otherestimators} for the explicit calculation for
other quadratic estimators. At leading order we have

\begin{eqnarray}
\left< \tilde E(\bfl_1)^\obs \tilde B(\bfl_2)^\obs\tilde
E(\bfl_1')^\obs\tilde B(\bfl_2')^\obs\right>_c
&=& (2\pi)^2 \delta_\dirac(\vecl_1+\vecl_2+\vecl_1'+\vecl_2')\times \nonumber \\
&&\quad \Bigg\{ C_{l_1}^{EE}C_{l_1'}^{EE}
\Bigg[C^\pp_{|\vecl_1+\vecl_2|}W_B(\vl_2,-\vl_1)W_B(\vl_2',-\vl_1')
 + C^\pp_{|\vecl_1+\vecl_2'|} W_B(\vl_2,-\vl_1')W_B(\vl_2',-\vl_1)\nonumber \\
&& \quad
+\,\sum^{P-distortion}_{S}C_{|{\vecl_1}+{\vecl_2}|}^{SS}W^{S}_{B}(\vl_2,-\vl_1)W^{S}_{B}(\vl_2',-\vl_1')\,+
\,\sum^{P-distortion}_{S}C_{|{\vecl_1}+{\vecl_2'}|}^{SS}W^{S}_{B}(\vl_2,-\vl_1')W^{S}_{B}(\vl_2',-\vl_1)\Big] \nonumber \\
&&\quad + \,C_{l_1}^{TE}C_{l_1'}^{TE}
\Big[\,\sum^{T-leakage}_{S}C_{|{\vecl_1}+{\vecl_2}|}^{SS}W^{S}_{B}(\vl_2,-\vl_1)W^{S}_{B}(\vl_2',-\vl_1')\,+ \nonumber \\
&&\quad\,\sum^{T-leakage}_{S}C_{|{\vecl_1}+{\vecl_2'}|}^{SS}W^{S}_{B}(\vl_2,-\vl_1')W^{S}_{B}(\vl_2',-\vl_1)\Bigg
]\Bigg\}
 \, , \label{eqn:trilens1}
 \end{eqnarray}
where we defined the lensing B-mode window function
 $W_B(\vl,\vl')\equiv \vl'\cdot(\vl-\vl')\sin
2(\varphi_{\vl}-\varphi_{\vl'})$ and $W^{S}_{B}(\vl_1,-\vl_1')$ is
the systematics window function for any of the 11 systematics
parameters. The formula for each of the systematic window functions
$W_{B}^S(\vl,\vl')$ can be found in Table~\ref{table:geometricd}.

\begin{figure}[]
\centering
\includegraphics[scale=1.2,clip] {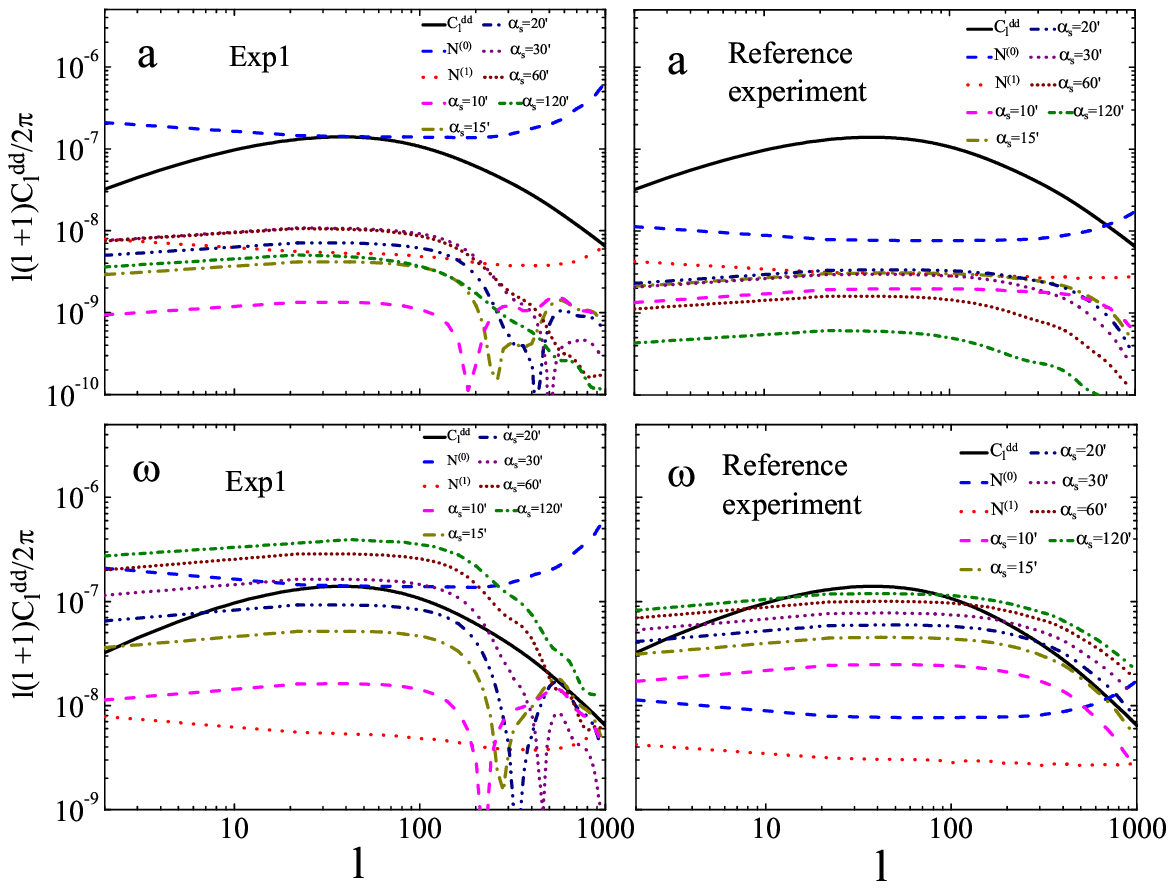}
\caption{{\it Upper panels}: Contamination from the calibration
systematics $a$ to the deflection angle power spectrum using the EB
estimator. The $\it rms$ fluctuation, $A_S$ is assumed to be 10$\%$.
The left and right panel are for Exp1 and reference experiment
respectively. In both the panels, the solid black, dashed blue, and
dot red curves show deflection angle power spectrum $C^{dd}(L)$,
Gaussian noise $N^{(0)}(L)$, and the first order non-Gaussian noise
$N^{(1)}(L)$. The remaining curves show the absolute value of the
systematic bias for various choices of coherence length $\alpha_s$,
starting from $\alpha_s=10'$ to $\alpha_s=120'$. {\it Lower panels}:
Same as the upper panels but for rotation systematics $\rot$.}
\label{fig:aandw}
\end{figure}

\begin{figure}[]
\centering
\includegraphics[scale=1.2,clip] {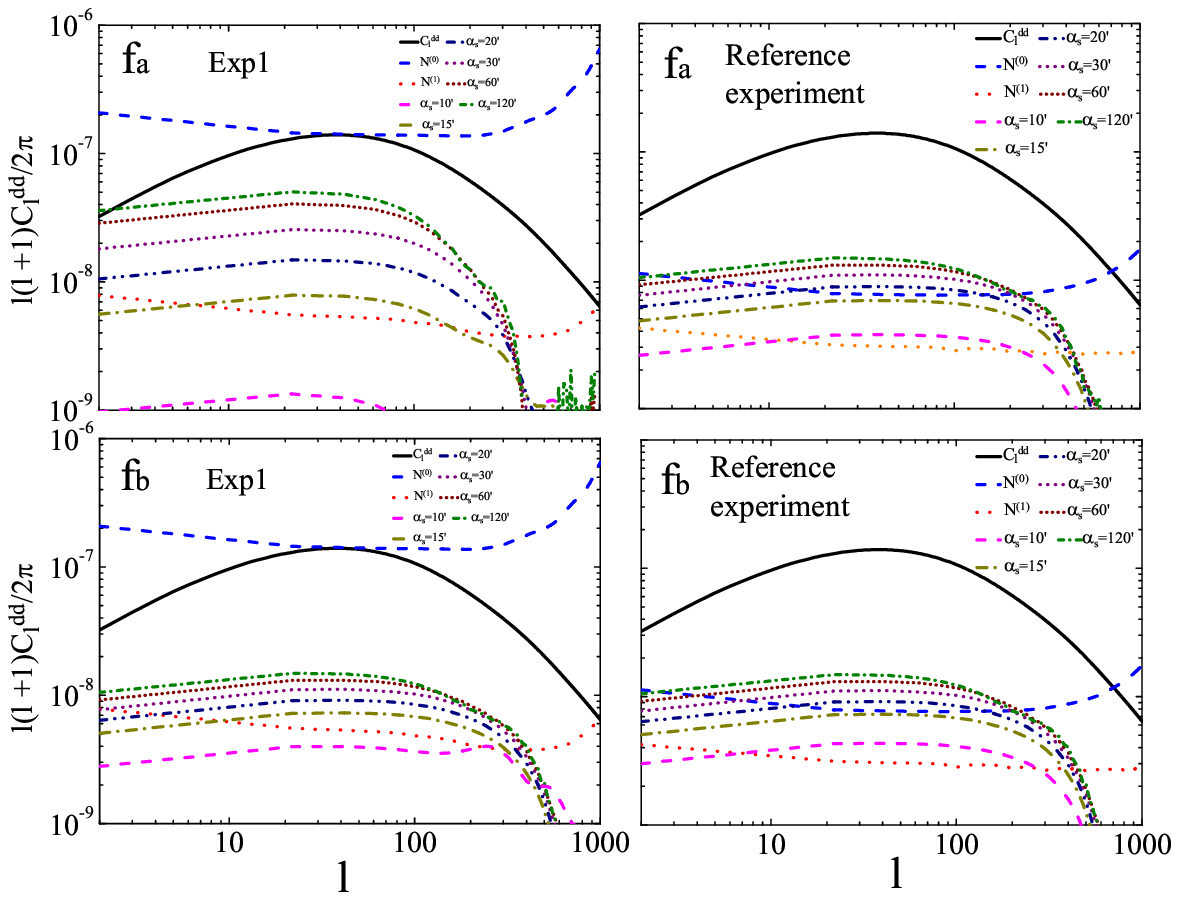}
\caption{{\it Upper panels}: Contamination from the spin-flip
systematics $f_a$ to the deflection angle power spectrum using the
EB estimator. The $\it rms$ fluctuation, $A_S$ is assumed to be
10$\%$. The left and right panel are for Exp1 and reference
experiment respectively. In both the panels, the solid black, dashed
blue, and dot red curves show deflection angle power spectrum
$C^{dd}(L)$, Gaussian noise $N^{(0)}(L)$, and the first order
non-Gaussian noise $N^{(1)}(L)$. The remaining curves show the
absolute value of the systematic bias for various choices of
coherence length $\alpha_s$, starting from $\alpha_s=10'$ to
$\alpha_s=120'$. {\it Lower panels}: Same as the upper panels but
for spin-flip systematics $f_b$.} \label{fig:faandfb}
\end{figure}

\begin{figure}[]
\centering
\includegraphics[scale=1.2,clip] {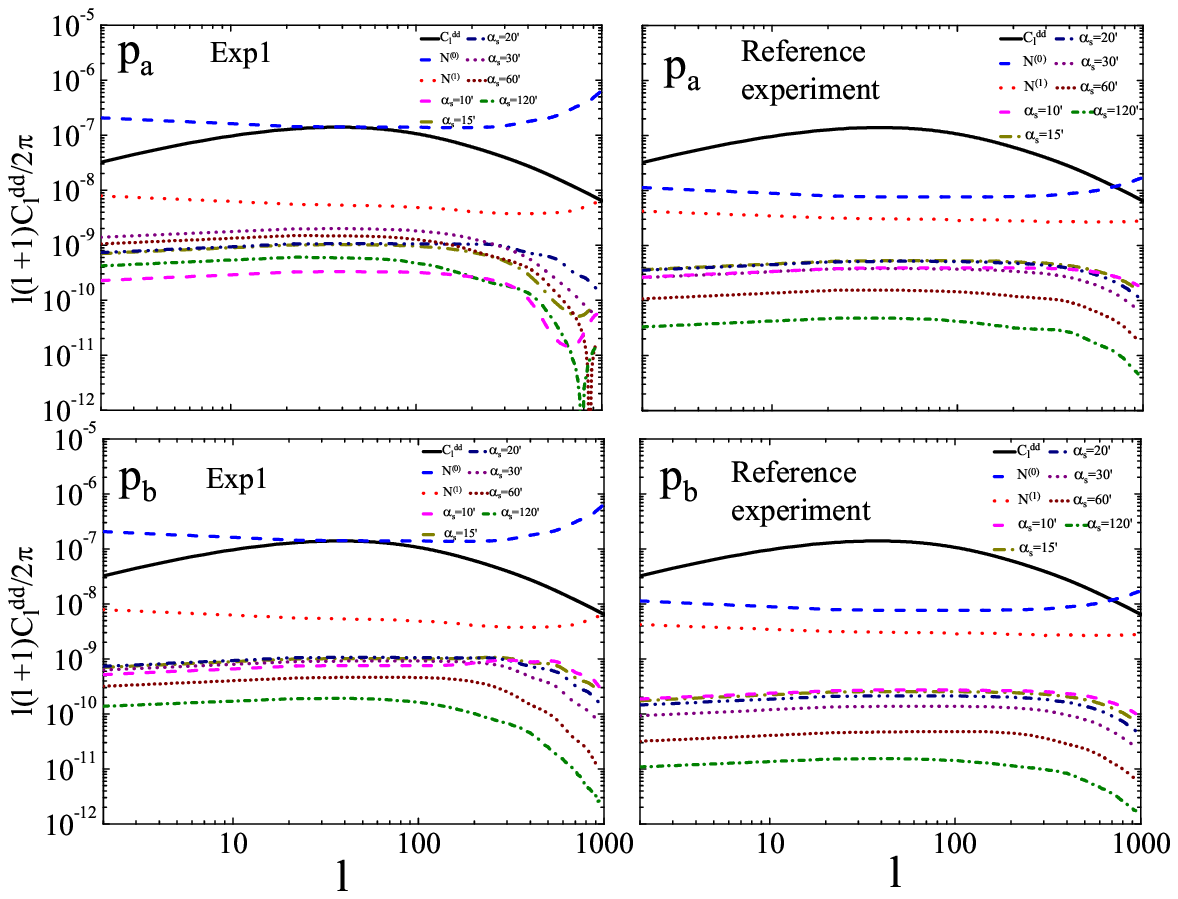}
\caption{{\it Upper panels}: Contamination from the pointing
systematics $p_a$ to the deflection angle power spectrum using the
EB estimator. The $\it rms$ fluctuation, $A_S$ is assumed to be
10$\%$. The left and right panel are for Exp1 and reference
experiment respectively. In both the panels, the solid black, dashed
blue, and dot red curves show deflection angle power spectrum
$C^{dd}(L)$, Gaussian noise $N^{(0)}(L)$, and the first order
non-Gaussian noise $N^{(1)}(L)$. The remaining curves show the
absolute value of the systematic bias for various choices of
coherence length $\alpha_s$, starting from $\alpha_s=10'$ to
$\alpha_s=120'$. {\it Lower panels}: Same as upper panels but for
the pointing systematics $p_b$.} \label{fig:panadpb}
\end{figure}

\begin{figure}[]
\centering
\includegraphics[scale=1.2,clip] {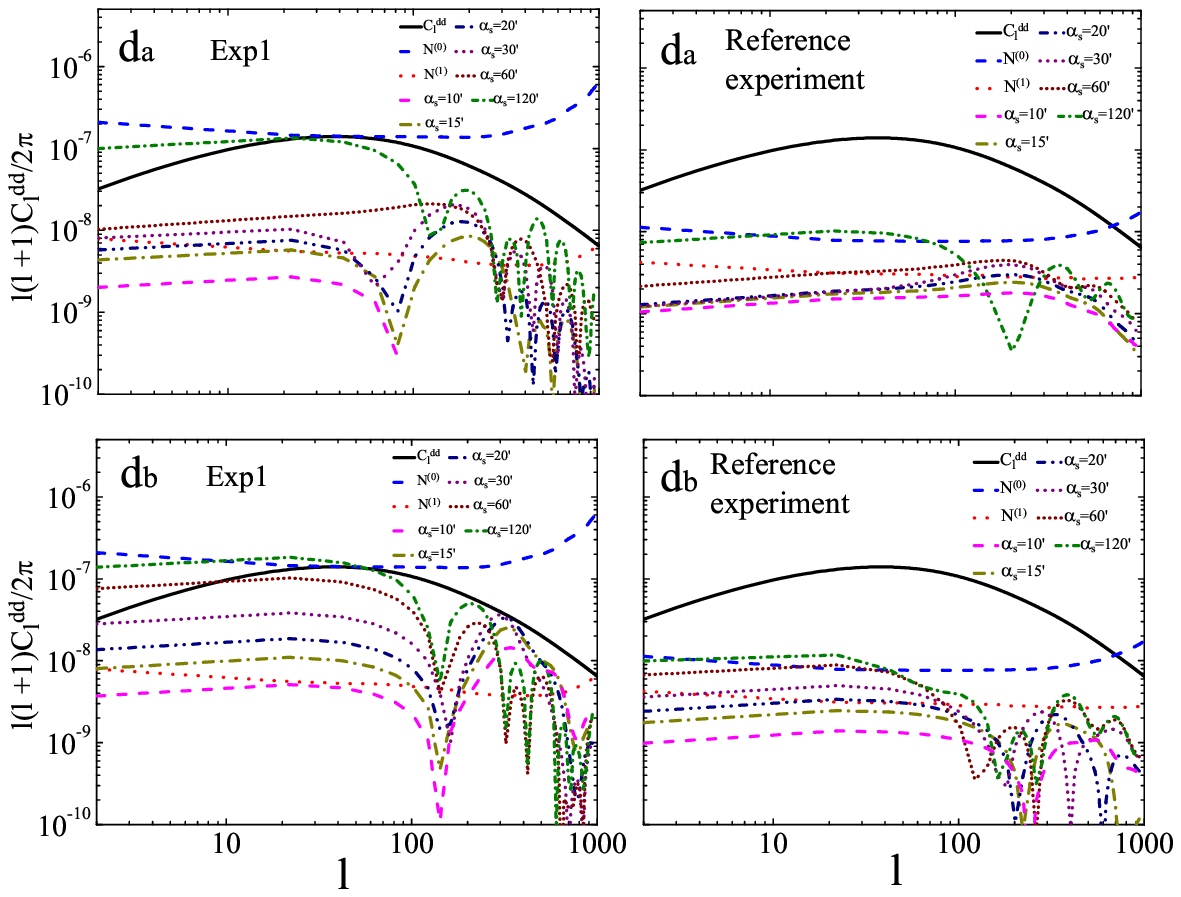}
\caption{{\it Upper panels}: Contamination from the  dipole leakage
$d_a$ to the deflection angle power spectrum using the EB estimator.
The $\it rms$ fluctuation, $A_S$ is assumed to be 10$\%$. The left
and right panel are for Exp1 and reference experiment respectively.
In both the panels, the solid black, dashed blue, and dot red curves
show deflection angle power spectrum $C^{dd}(L)$, Gaussian noise
$N^{(0)}(L)$, and the first order non-Gaussian noise $N^{(1)}(L)$.
The remaining curves show the absolute value of the systematic bias
for various choices of coherence length $\alpha_s$, starting from
$\alpha_s=10'$ to $\alpha_s=120'$. {\it Lower panels}: Same as upper
panels but for the dipole leakage $d_b$.} \label{fig:daanddb}
\end{figure}

\begin{figure}[]
\includegraphics[scale=1.2,clip] {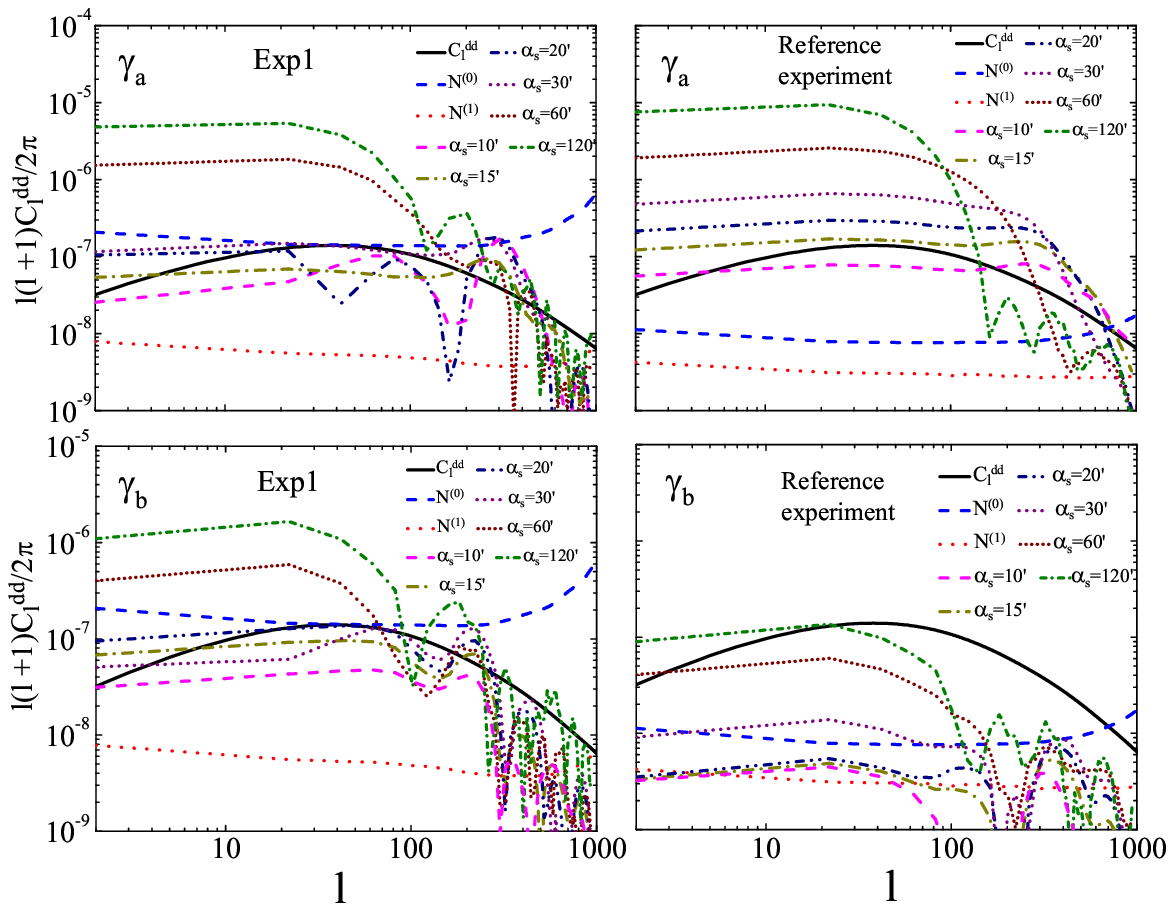}
\caption{{\it Upper panels}: Contamination from the monopole leakage
$\gamma_a$ to the deflection angle power spectrum using the EB
estimator. The $\it rms$ fluctuation, $A_S$ is assumed to be 10$\%$.
The left and right panel are for Exp1 and reference experiment
respectively. In both the panels, the solid black, dashed blue, and
dot red curves show deflection angle power spectrum $C^{dd}(L)$,
Gaussian noise $N^{(0)}(L)$, and the first order non-Gaussian noise
$N^{(1)}(L)$. The remaining curves show the absolute value of the
systematic bias for various choices of coherence length $\alpha_s$,
starting from $\alpha_s=10'$ to $\alpha_s=120'$. {\it Lower panels}:
Same as the upper panels but for the monopole leakage $\gamma_b$.}
\label{fig:raandrb}
\end{figure}

\begin{figure}[]
\centering
\includegraphics[scale=.9,clip] {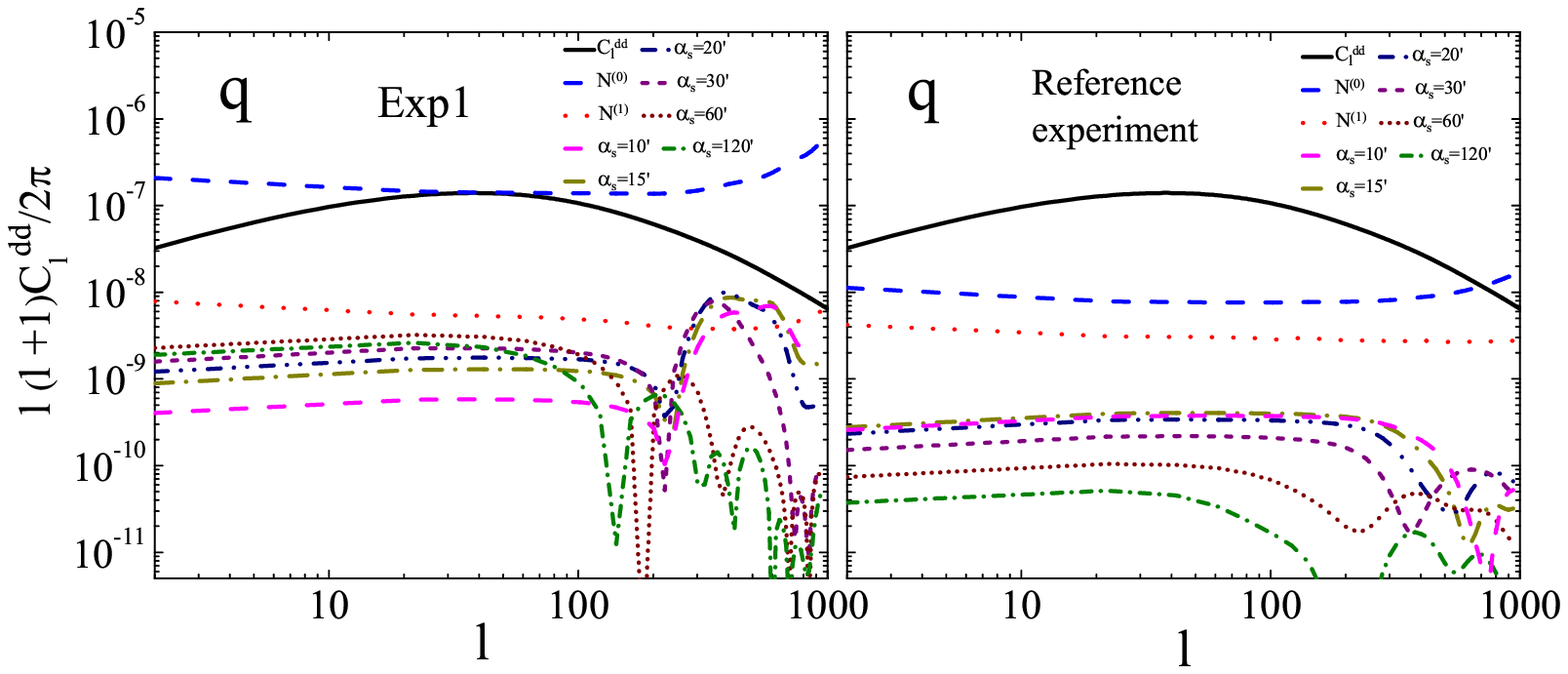}
\caption{Contamination from the quadrupole leakage $q$ to the
deflection angle power spectrum using the EB estimator. The $\it
rms$ fluctuation, $A_S$ is assumed to be 10$\%$. The left and right
panel are for Exp1 and reference experiment respectively. In both
the panels, the solid black, dashed blue, and dot red curves show
deflection angle power spectrum $C^{dd}(L)$, Gaussian noise
$N^{(0)}(L)$, and the first order non-Gaussian noise $N^{(1)}(L)$.
The remaining curves show the absolute value of the systematic bias
for various choices of coherence length $\alpha_s$, starting from
$\alpha_s=10'$ to $\alpha_s=120'$.} \label{fig:q}
\end{figure}

Different trispectrum can be constructed from combinations of the
temperature and polarization fields. We discuss other cases in the
Appendix~\ref{otherestimators} in order to calculate EE, TE and TB
estimators including contributions from systematics contamination.
The formulas shown here are readily generalized to the full sky. For
a discussion of the spherical generalization of the polarization
trispectra, see \cite{fullsky}. We have shown that E and B-modes are
mixed not only by weak lensing, but also by instrumental
systematics. Even if there is no lensing induced correlation,
certain kinds of systematics can give a non-zero contributions to
the trispectrum. Now we move on to construct quadratic lensing
estimators from E/B polarization modes, and we quantitatively show
how instrumental systematics contamination affects the lensing
reconstruction process. Again we take the EB estimator as an
example, and leave the discussions of other estimators to the
Appendix~\ref{otherestimators}.
The variance of the deflection angle power spectrum including systematic effects can be written as

\begin{eqnarray}
\Big\langle\Big\langle \langle \estEB(\bfL) \cdot \estEB(\bfLp)
\rangle_{\rm{CMB}} } \Big\rangle_{\rm{LSS}}\Big\rangle_{\rm{SYS} &=&
\frac{\normEB(L)}{L} \frac{\normEB(L')}{L'} \intl{1} \intl{1'}
\filtEB(\vecla, \veclb) \filtEB(\vecla', \veclb') \nonumber \\
&&\quad \times \left< \tilde E(\bfl_1)^\obs \tilde
B(\bfl_2)^\obs\tilde E(\bfl_1')^\obs\tilde B(\bfl_2')^\obs\right>
\nonumber \\
&=& (2 \pi)^2
\delta_\dirac(\bfL + \bfLp) \Bigg[ C^{dd}(L)+\noiseEB(L) + \noisePEB(L) +
\noiseSEB(L) + ...\Bigg] \,,
\label{eq:varianceEB}
\end{eqnarray}
where $\bfL=\vecla + \veclb$, and $C^{dd}(L)$ is the deflection
angle power spectrum. The terms $\noiseEB(L)$, $\noisePEB(L)$, and
$\noiseSEB(L)$ are the Gaussian noise, first order non-Gaussian
noise, and the first order systematics noise.  The Gaussian noise
contribution comes from the disconnected part of the four-point
function, while both the first-order non-Gaussian noise
$\noisePEB(L)$ and systematics noise $\noiseSEB(L)$ contribution
comes from the connected part. In deriving Eq. (\ref{eq:varianceEB})
we have used filters $F_{EB}$ which are given in Eq. (\ref{eq:F}),
and the trispectrum which is given in Eq. (\ref{eqn:trilens1}). The
ellipses stands for terms beyond first order in the systematics
power spectra or the deflection angle power spectrum. We note that
the Gaussian noise term also includes systematic effects implicitly
since instrumental systematics bias the measured power spectrum as
we have shown. The Gaussian and non-Gaussian noise terms in the
absence of systematic contribution have been previously reported in
~\cite{HuOkamoto} and ~\cite{Kesden}. The systematics noise term
$\noiseSEB(L)$ is new, for which the explicit form is given as
\begin{eqnarray}
\noiseSEB(L)&=&\frac{\normEB(L)}{L} \frac{\normEB(L')}{L'}\intl{1}
\intl{1'}
\filtEB(\vecla, \veclb) \filtEB(\vecla', \veclb')\nonumber \\
&&\quad \Bigg\{ C_{l_1}^{EE}C_{l_1'}^{EE} \Bigg [
\sum^{P-distortion}_{S}C_{|{\vecl_1}+{\vecl_2}|}^{SS}W^{S}_{B}(\vl_2,-\vl_1)W^{S}_{B}(\vl_2',-\vl_1')\,+
\,\sum^{P-distortion}_{S}C_{|{\vecl_1}+{\vecl_2'}|}^{SS}W^{S}_{B}(\vl_2,-\vl_1')W^{S}_{B}(\vl_2',-\vl_1)\Bigg ] \nonumber \\
&&\quad + \,C_{l_1}^{TE}C_{l_1'}^{TE} \Bigg
[\,\sum^{T-leakage}_{S}C_{|{\vecl_1}+{\vecl_2}|}^{SS}W^{S}_{B}(\vl_2,-\vl_1)W^{S}_{B}(\vl_2',-\vl_1')\,+
\sum^{T-leakage}_{S}C_{|{\vecl_1}+{\vecl_2'}|}^{SS}W^{S}_{B}(\vl_2,-\vl_1')W^{S}_{B}(\vl_2',-\vl_1)\Bigg
]\Bigg\}\,. \label{eq:Ns}
\end{eqnarray}
Eq. (\ref{eq:Ns}) is our main result for the systematics contamination on lensing power reconstruction using the EB estimator.
We will use this equation to numerically compute the systematic-induced bias for the 11 systematic parameters.
The results are shown in Figures (\ref{fig:aandw}) to (\ref{fig:q}) and are discussed
in next section\footnote{As a note, for each instrumental systematics parameter $S$, for both the polarization
 distortion and temperature leakage, there are two terms which
contribute to the final results. The terms proportional
$C^{SS}_{|{\vecl_1}+{\vecl_2}|}$ is generally much smaller than the
terms proportional to $C^{SS}_{|\vecl_1+\vecl_2'|}$  but for
non-Gaussian noise $N^{(1)}$~\cite{Kesden}, the terms proportional
to $C^\pp_{|{\vecl_1}+{\vecl_2}|}$ are about an order of magnitude
larger than other two terms proportional to
$C^\pp_{|\vecla+\vecla'|}$ and $C^\pp_{|\vecla+\veclb'|}$.}.

\begin{table}
\begin{tabular}{||c||cc|cc|cc||cc|cc|cc||}
\hline \hline

  &  \multicolumn{6}{|c||}{\sl Exp1 } &  \multicolumn{6}{|c||}{\sl Reference} \\
\cline{2-13} \bf {Type} &  \multicolumn{2}{|c|}{Lensing } &
\multicolumn{4}{|c||}{B-modes} & \multicolumn{2}{|c|}{Lensing} &
          \multicolumn{4}{|c||}{B-modes}\\
\cline{2-13} &  \multicolumn{2}{|c|}{ $\ell$ = 40} &
\multicolumn{2}{|c|}{3.0$\times10^{16}$ Gev} &
\multicolumn{2}{|c||}{1.0$\times10^{16}$ GeV} & \multicolumn{2}{|c|}{
$\ell$ = 40} & \multicolumn{2}{|c|}{3.0$\times10^{16}$ GeV} &
          \multicolumn{2}{|c||}{1.0$\times10^{16}$ GeV}\\
\cline{2-13}
& $\alpha_{s}=10^\prime$ & $\alpha_{s}=120^\prime$ & $\alpha_{s}=10^\prime$ & $\alpha_{s}=120^\prime$ & $\alpha_{s}=10^\prime$ & $\alpha_{s}=120^\prime$& $\alpha_{s}=10^\prime$ & $\alpha_{s}=120^\prime$ & $\alpha_{s}=10^\prime$ & $\alpha_{s}=120^\prime$ & $\alpha_{s}=10^\prime$ & $\alpha_{s}=120^\prime$\\
\hline Calibration $a$
&  1.04  &    0.55 &  0.549      &  0.468      &   0.061  &    0.052 &  0.86  &   1.56 &   0.486       &  0.468       &   0.054 &    0.052 \\
Rotation $w$
&  0.30  &    0.061 & 0.27       &  0.207      &    0.030  &    0.023 &  0.24  &   0.11 &  0.243      &  0.198      &     0.027 &    0.022 \\
Pointing $p_a$
&  2.10  &    1.57 &  8.55      &  6.12      &    0.95  &    0.68 &  1.93  &   5.55 &  15.3      &  12.24      &     1.70 &    1.36 \\
Pointing $p_b$
&  1.39  &    2.76 &  1.08      &   6.57     &    0.12  &    0.73 &  2.31  &   9.74 &  1.71      &  13.14      &    0.19 &    1.46 \\
Flip $f_a$
& 1.08  &    0.17 &  0.549      &   0.441     &    0.061  &    0.049 &  0.62  &   0.31 &  0.486      &  0.432      &    0.054 &    0.048 \\
Flip $f_b$
&  0.61  &    0.17 &   0.531     &  0.360      &    0.059  &    0.040 &  0.58  &   0.31 &  0.477      &  0.36      &     0.053 &    0.040 \\
Monopole $\gamma_a$
&  0.114  &    0.024 &  0.021      &  0.0058      &    0.0023  &    0.00064 &  0.13  &   0.013 &  0.0207       &  0.0058      &     0.0023 &    0.00064 \\
Monopole $\gamma_b$
&  0.114  &    0.036 &  0.014      &  0.0034      &    0.0016  &    0.00038 &  0.64  &   0.12 &   0.0144     &  0.0034      &     0.0016 &    0.00038 \\
Dipole $d_a$
&  0.82  &    0.11 &   0.085     &  0.060      &    0.0094  &    0.0067 &  0.97  &   0.39 &  0.153      &   0.117     &     0.017 &    0.013 \\
Dipole $d_b$
&  0.55  &    0.092 &  0.85      &  0.063      &    0.0094  &    0.0070 &  1.02  &   0.33 &  0.153      &  0.126      &     0.017 &    0.014 \\
Quadrupole $q$
&  1.58  &    0.78 &   0.162     &  0.558      &    0.018  &    0.062 &  2.38  &   5.47 &  0.495      &  2.25      &    0.055 &    0.25 \\

\hline
\end{tabular}
\caption{Systematic contamination for lensing and B-mode detection
for the two experimental setups (Exp1 and reference), for two
choices of coherence length $\alpha_{s}=10'$ and  $\alpha_s=120'$.
We scale the $\it rms$ amplitude of the systematics field $A_s$
 to the same level as the maximum of the signal (deflection angle power spectrum at $\ell=40$ and B-mode signal for a fiducial energy scale at $\ell=90$).
 First column mentions the type of systematic parameter in consideration. Within the lensing columns we show the maximum $A_s$ (for $\alpha_s=10'$ and $\alpha_s=120'$)
 for bias to not exceed
the lensing power spectrum. For B-modes columns we show the maximum
$A_s$ (for $\alpha_s=10'$ and $\alpha_s=120'$) for bias to not
exceed the B-mode signal for two choices of inflationary energy
scale $E=3.0\times10^{16}$ GeV and $E=1.0\times10^{16}$ GeV
respectively. } \label{table:syst}
\end{table}

\begin{table*}
\begin{center}
\begin{tabular}{||c|cc||cc||}
\hline\hline

  &  \multicolumn{2}{|c||}{\sl Exp1 } &  \multicolumn{2}{|c||}{\sl Reference} \\
  \cline{2-5}
\bf {Type}          &  \multicolumn{2}{|c||}{$E_{crit.}$(GeV) } &  \multicolumn{2}{|c||}{$E_{crit.}$(GeV)} \\
\cline{2-5}
          & $\alpha_{s}=10^\prime$ & $\alpha_{s}=120^\prime$ & $\alpha_{s}=10^\prime$ & $\alpha_{s}=120^\prime$ \\
\hline Calibration $a$
&  4.13  &    3.25 &   3.99  &    8.83 \\
Rotation $w$
&  3.16  &  1.63 &    2.98  &    2.24   \\
Pointing $p_a$
&  1.49  &  1.52 &    1.07  &     2.02 \\
Pointing $p_b$
&  3.40  &  1.94 &    3.49  & 2.58  \\
Flip $f_a$
&  4.21  &  1.86 &    3.34  &     2.54 \\
Flip $f_b$
&  3.22  &  2.06 &      3.31  &     2.78  \\
Monopole $\gamma_a$
&  7.80  &  5.00 &   7.80  &     4.33  \\
Monopole $\gamma_b$
&  8.29  &  9.73 &     20.00 &     17.77  \\
Dipole $d_a$
&  9.34  &  4.05 &      7.55 &     5.48  \\
Dipole $d_b$
&  7.51 &      3.57 &   7.78 &     5.00 \\
Quadrupole $q$
&  9.40 &      3.48 &   6.02  &     6.42  \\
\hline
\end{tabular}
\end{center}
\caption{We show the critical inflation energy scale $E_{crit.}$ for each systematic parameter, for two choices for coherence length ($\alpha_s=10'$ and $\alpha_s=120'$) and for two experimental setups (Exp1 and reference).  First column mentions the type of systematic parameter in consideration.
$E_{crit}$ of a given systematic parameter is the inflationary energy for which the systematic contamination (from that parameter) is equal to both the maximum B-modes signal (corresponding to $\ell=90$) and the lensing signal (corresponds to $\ell=40$). Hence for a given systematic parameter if the requirement for detecting B-modes with energy scale $E \le E_{crit.}$ have been met, then lensing extraction requirement are already met. Similarly if a given systematic parameter sets the detectable B-modes with energy scale $E \ge E_{crit.}$, then lensing extraction requirement provide more stringent constraint on that systematic parameter. } \label{table:Ecrit}
\end{table*}

\section{Results}
\label{results}

Figures \ref{fig:aandw} to \ref{fig:q} and Tables~\ref{table:syst}
$\&$ \ref{table:Ecrit} summarize our main findings. We have focused
on the systematics-induced bias for the $EB$ estimator (results for
the other estimators are provided in the
Appendix~\ref{otherestimators}) because it has the highest Gaussian
signal-to-noise ratio for reconstructing the projected matter power
spectrum (See Fig.~\ref{fig:gaussian} or~\cite{HuOkamoto}).
Figures~\ref{fig:aandw}-\ref{fig:q} show the contamination
introduced by different systematic effects (the term $\noiseSEB(L)$
in Eq. (\ref{eq:Ns})) in the deflection angle power spectrum
reconstruction. We have assumed the rms fluctuation of the
systematics fields to be $10\%$, and varied coherence length
starting from minimum $\alpha_s=10'$ to the maximum $\alpha_s=120'$.
For comparison, we also show the level of Gaussian noise and first
order non-Gaussian noise of the EB estimator, the terms
$\noiseEB(L)$ and $\noisePEB(L)$ respectively in Eq. (\ref{eq:Ns}).
Systematics-induced bias generally increases with the increase in
coherence length. For all the coherence lengths, the bias
$\noiseSEB(L)$ is fairly constant on large scales, similar to the
feature of the Gaussian noise and the first order non-Gaussian noise
of the estimator, however on small scales ($\ell \gtrsim 200$), the
systematic contamination are not constant, and for some systematics,
they oscillate between positive and negative values. The fluctuating
features are due to the oscillation of Gaussian noise at high $\ell$
and the CMB power spectra included in the systematic calculations,
smoothed and filtered by certain window functions. The negative
systematic contributions at certain multiple range for some
parameters are caused by combining the effects from window
functions, which sometimes give negative, and TE cross correlation
which enters the calculation of monopole leakage, dipole leakage,
and quadrupole leakage from temperature field.

 In Table~\ref{table:syst}
we calculate the maximum {\it rms} amplitude $A_s$ required for each
systematic parameters to keep its own contamination lower than the
deflection angle power spectrum. Lensing extraction requires the
control of systematic {\it rms} fluctuations at levels depending on
the type of systematics. Among all the 11 systematic parameters,
rotation $\rot$, and monopole leakage $\gamma_a \& \gamma_b$ place
the most stringent requirements, while pointing error $p_a \& p_b$,
quadrupole leakage $q$, and calibration $a$ are among the least
demanding. Also with respect to sensitivity to the coherence length,
some systematic parameters such as rotation $\rot$ and monopole
leakage $\gamma_a \& \gamma_b$ are very sensitive (as shown in
Figure (\ref{fig:aandw}) and (\ref{fig:daanddb})), while some are
not very sensitive.

For comparison we also show the systematics requirements for
primordial B-modes detection. For B-modes we consider two
inflationary energy scales $E=1.0\times10^{16}$ GeV and
$E=3.0\times10^{16}$ GeV. In addition we consider two choices of
coherence length $\alpha_s=10'$ and $\alpha_s=120'$, and two
experimental setups Exp1 and reference experiment.
 Since the systematic requirements from B-modes detection is
 sensitive to the inflationary scale in question, it is useful to define an energy scale $E_{crit}$ below which lensing reconstruction is safe and above which lensing sets the systematic requirements. We define the critical energy
$E_{crit}$ of a given systematic parameters as the inflationary energy for which the systematic requirements to extract the B-modes signal is equal to the requirement to extract the lensing signal. Hence for a given systematic parameter, if the requirement for detecting B-modes with energy scale $E < E_{crit.}$ have been met, then lensing extraction requirements are already met. Similarly if a given systematic parameter sets the detectable B-modes with energy scale $E > E_{crit.}$, then lensing extraction requirements provide more stringent constraints on that systematic parameter. We show the critical energy $E_{crit}.$ in Table \ref{table:Ecrit} for all the 11 systematic parameters for two choices of coherence length ($\alpha_s=10'$ and $\alpha_s=120'$) and for both the experimental setups (Exp1 and reference).

For both the Exp1 and reference experiment, lensing reconstruction
is safe once the experimental requirements to detect B-modes with
inflationary scale $E_i = 1.0\times10^{16}$ GeV are met. The
requirements from lensing extraction are about 1-2 orders of
magnitude less stringent than the requirements to measure the
primordial B-modes with inflationary energy scale of
$1.0\times10^{16}$ GeV. This means that once the experiment
satisfies the requirements of systematic control to detect
primordial B-modes, it is safe to use such an experiment to
reconstruct the lensing potential power spectrum without extra
effort on improving the instrumental systematic control. On the
other hand the requirements for lensing reconstruction are
comparable or even more stringent for some systematic parameters
than the requirements to detect primordial B-modes with inflationary
scale $E_i = 3.0\times10^{16}$ GeV.





\section{Conclusion}
\label{conclusion}

We illustrate the effects of instrumental systematics on the
reconstruction of the projected matter power spectrum from CMB
gravitational lensing. We consider seven types of effects which are
related to known instrumental systematics: calibration, rotation,
pointing, spin-flip, monopole leakage, dipole leakage and quadrupole
leakage. These effects can be parametrized by 11 distortion fields.
Each of these systematic effects can mimic the effective projected
matter power spectrum and hence contaminate the lensing
reconstruction.  We assume a Gaussian distribution for each
parameter. We have modeled the fluctuations in the instrumental
contamination fields with a coherence length ($\alpha_s$) and an
$\it rms$ amplitude ($A_s$), as defined in Eq. (\ref{eq:coh}) and
(\ref{eq:rms}), respectively. Rotation systematics $\rot$ and
monopole leakage $\gamma_a \& \gamma_b$ are among the most dangerous
ones.

It is important to know how the systematic effects propagate
to the lensing potential reconstruction with good precision, in order to reliably
reconstruct the lensing potential. Without well understood and calibrated
instrumental systematics, the extraction of cosmological parameters from the statistical properties of lensing potential would also be
biased. Systematics in the lensing reconstruction are also one of the key concerns for detecting
primordial gravitational waves by removing the contribution properly
from lensing-reduced B-modes. A faithful detection of the primordial gravitational wave signal
largely depends on how clean the delensing process has been.


\acknowledgments
APSY thanks Daniel Baumann, Ben Wandelt, Scott Kruger, Oliver
Zahn, and Jaiyul Yoo for useful discussions and help during the
project. We especially thank Daniel Baumann for reading the
manuscript and providing useful feedback.

{}

\appendix

\section{Instrumental systematics for the TT estimator}
\label{appentemsys}

In this appendix, we show detailed calculation of instrumental
systematic contamination on the TT estimator.  The calculation is
similar to the systematic contamination analysis for other
estimators involving the polarization fields (discussed in
Appendix~\ref{appenpolasys}).

We first expand the observed temperature $\tilde\cmb^{obs}(\bn)$
field, given by in Eq.~(\ref{E:expandsys}) as
\begin{eqnarray}
\label{E:RealTay}
\tilde\cmb^{obs}(\bn) & = &  [1+a(\bn)]\cmb[\bn + \nabla\len(\bn)] \nonumber\\
        & \approx &
\cmb(\bn) + \nabla_a \len(\bn) \nabla^a \cmb(\bn) + \frac{1}{2}
\nabla_a \len(\bn) \nabla_b \len(\bn) \nabla^{a}\nabla^{b} \cmb(\bn)
+ \ldots\nonumber\\
 &&\quad
+ a(\bn)\times[\cmb(\bn) + \nabla_a \len(\bn) \nabla^a \cmb(\bn) +
\frac{1}{2} \nabla_a \len(\bn) \nabla_b \len(\bn)
\nabla^{a}\nabla^{b} \cmb(\bn)
+ \ldots\, ],
\end{eqnarray}
where $a(\bn)$ is the calibration parameter. Taking the Fourier
transform of the lensed map with systematics $\tilde
\cmb^{obs}(\bn)$ under the flat-sky approximation,
\begin{eqnarray}
\tilde \cmb^\obs(\bfl)
&=& \int d \bn\, [1+a(\bn)]\tilde \cmb(\bn) e^{-i \bfl \cdot \bn} \nonumber\\
&=& \cmb(\bfl) - \intlnp \cmb(\bflp) L(\bfl,\bflp)\,,
\label{E:thetal}
\end{eqnarray}
where
\begin{eqnarray}
\label{E:lfactor}
L(\bfl,\bflp) &\equiv& \len(\bfl-\bflp)\left[(\bfl-\bflp) \cdot \bflp \right] +\frac{1}{2} \intlnpp \len(\bflpp)\times\len(\bfl - \bflp - \bflpp) \, (\bflpp \cdot
\bflp) \left[ (\bflpp + \bflp - \bfl)\cdot\bflp \right] + \ldots \nonumber\\
&&\quad
-a(\bfl-\bflp) + \intlnpp
a(\bflpp)\times[\bflp\cdot(\bfl-\bflp-\bflpp)]\,
\len(\bfl-\bflp-\bflpp) \nonumber\\
&&\quad
+\frac{1}{2} \intlnpp \intlnppp
a(\bflppp)\times\len(\bflpp)\times\len(\bfl - \bflp -
\bflpp-\bflppp)\, (\bflpp \cdot \bflp)\left[ (\bflppp+ \bflpp +
\bflp - \bfl) \cdot \bflp \right] + \ldots \,.
\end{eqnarray}
Using the above equations one can compute the CMB temperature power
spectrum $C^{TT}_{l}$ in presence of systematic contamination, the
result is given in Eq. (\ref{E:lenpower}).

Analysis of the deflection angle power spectrum $C_l^{dd}$ involves
trispectrum calculation. Below we show the details of the temperature trispectrum including the
systematic contribution,
\begin{eqnarray} \label{E:4pt,CMB}
&&\langle \cmb^{obs}(\vecla) \cmb^{obs}(\veclb)\cmb^{obs}(\vecla')
\cmb^{obs}\veclb') \rangle_{\rm{CMB}} = \nonumber \\
&&\quad \Biggl\{ \left(
C_{l_1}^{\cmb\cmb} + C_{l_1}^{\cmb\cmb\n} \right) (2 \pi)^2
\delta_\dirac(\bfL) + a(\bfL)\left( C_{l_1}^{\cmb\cmb}+
C_{l_2}^{\cmb\cmb}\right) + \len(\bfL)
\ffact(\vecla, \veclb)
  - \intl{1'} a(\vecla')(\bfL-\vecla')\len
(\bfL - \vecla')\veclb C_{l_2}^{\cmb\cmb}\nonumber \\
&& \quad - \intl{2'} a(\veclb')(\bfL-\veclb')\len (\bfL -
\veclb')\vecla C_{l_1}^{\cmb\cmb}  \nonumber 
  + \frac{1}{2} \intl{1'} \len(\vecla') \len(\bfL - \vecla')
\Bigl\{ C_{l_1}^{\cmb\cmb} (\vecla \cdot \vecla') \bigl[ \vecla
\cdot (\bfL - \vecla') \bigr] + C_{l_2}^{\cmb\cmb} (\veclb \cdot
\vecla') \bigl[
\veclb \cdot (\bfL - \vecla') \bigr] \Bigr\} \nonumber \\&& \quad
+\intl{1'}a(\vecla-\vecla')a(\veclb+\vecla')C_{l_1^{\prime}}^{\cmb\cmb}
 \nonumber 
- \intl{1'} C_{l_1^{\prime}}^{\cmb\cmb} \len(\vecla -
\vecla') \len(\veclb + \vecla') \bigl[ \vecla' \cdot (\vecla -
\vecla')
\bigr]\times \bigl[ \vecla' \cdot (\veclb + \vecla') \bigr]  \nonumber \\
&& \quad - \intl{2'} C_{l_2^{\prime}}^{\cmb\cmb} \len(\veclb -
\veclb') a(\vecla + \veclb') \bigl[ \veclb' \cdot (\veclb - \veclb')
\bigr] \nonumber 
- \intl{1'} C_{l_1^{\prime}}^{\cmb\cmb} \len(\vecla -
\vecla') a(\veclb + \vecla') \bigl[ \vecla' \cdot (\vecla - \vecla')
\bigr]\Biggl\}\nonumber \\
&& \times  \Biggl\{ \left( C_{l_1^{\prime}}^{\cmb\cmb} +
C_{l_1^{\prime}}^{\cmb\cmb\n} \right) (2 \pi)^2 \delta_\dirac(\bfL)
+ a(\bfL)\left( C_{l_1^{\prime}}^{\cmb\cmb}+
C_{l_2^{\prime}}^{\cmb\cmb}\right) + \len(\bfL) \ffact(\vecla',
\veclb')  \nonumber 
- \intln a(\vecl)(\bfL-\vecl)\len
(\bfL - \vecl)\veclb' C_{l_2^{\prime}}^{\cmb\cmb}\nonumber \\
&& \quad - \intln a(\vecl)(\bfL-\vecl)\len (\bfL - \vecl)\vecla'
C_{l_1^{\prime}}^{\cmb\cmb}   \nonumber 
+ \frac{1}{2} \intln \len(\vecl) \len(\bfL - \vecl) \Bigl\{
C_{l_1^{\prime}}^{\cmb\cmb} (\vecla'\cdot \vecl) \bigl[ \vecla'
\cdot (\bfL - \vecl) \bigr] +C_{l_2^{\prime}}^{\cmb\cmb} (\veclb'
\cdot \vecl) \bigl[
\veclb' \cdot (\bfL - \vecl) \bigr] \Bigr\} \nonumber \\
&&\quad + \intln a(\vecla'-\vecl)a(\veclb'+\vecl)C_{l}^{\cmb\cmb}  \nonumber 
- \intln C_{l}^{\cmb\cmb} \len(\vecla' - \vecl)
\len(\veclb' + \vecl) \bigl[ \vecl \cdot(\vecla' - \vecl) \bigr]
\times \bigl[ \vecl \cdot (\veclb' + \vecl) \bigr] \nonumber \\
&& \quad -\intln C_{l}^{\cmb\cmb} \len(\veclb' - \vecl) a(\vecla' +
\vecl) \bigl[ \vecl \cdot (\veclb' - \vecl) \bigr] - \intln
C_{l}^{\cmb\cmb} \len(\vecla' - \vecl) a(\veclb' + \vecl) \bigl[
\vecl \cdot (\vecla' - \vecl) \bigr]\Biggl\}  +  Perm.
\end{eqnarray}

In practice, we do not know the large-scale structure between us and
the last-scattering surface, we must average over different
realizations of the large-scale structure (denoted by $\langle \quad
\rangle_{\rm{LSS}}$) and also over different realizations of the
systematic fields (denoted by $\langle \quad\rangle_{\rm{SYS}}$) to
obtain the observed deflection angle power spectrum. The deflection
field power spectrum can be estimated by the TT estimator as
\begin{eqnarray} \label{E:estvar}
\Big\langle\Big\langle \langle \est(\bfL) \cdot \est(\bfLp)
\rangle_{\rm{CMB}}\Big\rangle_{\rm{LSS}}\Big\rangle_{\rm{SYS}}
&\equiv& \langle \est(\bfL) \cdot \est(\bfLp) \rangle \nonumber \\
&=& (\bfL \cdot \bfLp) \frac{\norm(L)}{L^2}
\frac{\norm(L^{\prime})}{L^{\prime2}} \nonumber \\
&&  \times \intl{1} \intl{1'} \Big\langle\Big\langle \langle
\cmb^\obs(\vecla) \cmb^\obs(\veclb) \cmb^\obs(\vecla')
\cmb^\obs(\veclb')
\rangle_{\rm{CMB}}\Big\rangle_{\rm{LSS}}\Big\rangle_{\rm{SYS}}\nonumber \\
&&  \times \filt(\vecla, \veclb) \filt(\vecla', \veclb').
\end{eqnarray}
Only the connected part of the trispectrum comes into the deflection
angle power spectrum $C^{dd}_{\ell}$. Keeping the first
order of $C_{l}^{\pp}$ and $C_{l}^{aa}$, the results can be
simplified as
\begin{eqnarray}
\langle \cmb^{obs}(\vecla) \cmb^{obs}(\veclb) \cmb^{obs}(\vecla')
\cmb^{obs}(\veclb') \rangle_{\rm{CMB},conn.} &=& \Big<\intl{1'}
\cmb(\vecla')
\len(\vecla-\vecla') \, [(\vecla - \vecla')\cdot \vecla']\nonumber \\
&& \quad \times \intl{2'} \cmb(\veclb') \len(\veclb-\veclb') \,
[(\veclb - \veclb') \cdot \veclb']\,
\cmb(\vecla') \cmb(\veclb') \Big> \nonumber \\
&& \quad + \Big<\intl{1'}\cmb(\vecla') a(\vecla -\vecla')\ \intl{2'}
\cmb(\veclb') a(\veclb -\veclb')
\cmb(\vecla')  \cmb(\veclb') \Big> \nonumber \\
&=& C_{l_1'}^\cmb C_{l_2'}^\cmb \Big< \len(\vecla+\vecla')
\len(\veclb+\veclb') \Big> \, [(\vecla + \vecla') \cdot \vecla'] \,
[(\veclb + \veclb') \cdot \veclb'] \nonumber \\
&& \quad + C_{l_1'}^\cmb C_{l_2'}^\cmb \Big< a(\vecla+\vecla')
a(\veclb+\veclb') \Big> \,  \nonumber \\
&& \quad + C_{l_1'}^\cmb C_{l_2'}^\cmb \Big< \len(\vecla+\veclb')
\len(\veclb+\vecla') \Big> \, [(\vecla + \veclb') \cdot \veclb'] \,
[(\veclb + \vecla') \cdot \vecla'] \nonumber \\
&& \quad  + C_{l_1'}^\cmb C_{l_2'}^\cmb \Big< a(\vecla+\veclb')
a(\veclb+\vecla') \Big> \, + Perm.\, ,
\end{eqnarray}
The terms that are linear in the lensing potential $\len(\vecl)$ and the
temperature systematics $a(\vecl)$ vanish. Further
averaging over the lensing field and the systematics field, one can get
Eq.~(\ref{E:temsys}) which gives the measured lensing induced CMB
temperature trispectrum in the presence of instrumental systematic
contaminations.

\section{Polarization Systematics Contamination}
\label{appenpolasys}

In this Appendix we calculate the window functions (given in
Table~\ref{table:geometricd}) for each instrumental systematic. We
show how the systematic contamination comes in the analysis
inevitably. Again the calculation is done for small deflection
angles (weak lensing).

We will first do the calculation for the polarization transfer
systematics. The Taylor expansion of two Stokes polarization
parameters including the systematic parameters can be written as:

\begin{eqnarray}
\label{E:Polamea} [\tQ \pm i \tU]^{obs}(\bn) & = & [1+\calb \pm i
2\rot](\bn)[Q \pm i U][\bn + \nabla\len(\bn)]
+ [f_1 \pm i f_2](\bn)[Q \mp i U][\bn + \nabla\len(\bn)]+ [\gamma_1 \pm i \gamma_2](\bn) \cmb[\bn + \nabla\len(\bn)]\nonumber\\
&=& [Q \pm i U](\bn) + \nabla_a \len(\bn) \nabla^a [Q \pm i U](\bn)
+ \frac{1}{2}\nabla_a \len(\bn) \nabla_b \len(\bn)
\nabla^{a}\nabla^{b}[Q \pm i U](\bn)]+ \ldots \nonumber\\
&& + [\calb \pm i 2 \rot](\bn)[\,[Q \pm i U](\bn) + \nabla_a
\len(\bn) \nabla^a [Q \pm i U](\bn) + \frac{1}{2}\nabla_a \len(\bn)
\nabla_b \len(\bn) \nabla^{a}\nabla^{b}[Q \pm i U](\bn)+ \ldots]\nonumber\\
&& + [f_1 \pm i f_2](\bn)[\,[Q \mp i U](\bn) + \nabla_a \len(\bn)
\nabla^a [Q \mp i U](\bn) + \frac{1}{2}\nabla_a \len(\bn) \nabla_b
\len(\bn) \nabla^{a}\nabla^{b}[Q \mp i U](\bn)+ \ldots]\nonumber]\\
&& + [\gamma_1 \pm i \gamma_2](\bn) [\cmb(\bn) + \nabla_a \len(\bn)
\nabla^a \cmb(\bn) + \frac{1}{2}\nabla_a \len(\bn) \nabla_b
\len(\bn) \nabla^{a}\nabla^{b}\cmb(\bn)+ \ldots]\,.
\end{eqnarray}
Performing the harmonic transformation we get
\begin{eqnarray}
[\tilde E(\vl)\pm i \tilde B(\vl)]^{obs} &=& \int d \bn\, [\tQ \pm i \tU]^{obs}(\bn) e^{\mp2i\varphi_\vl} e^{-i \bfl \cdot \bn} \nonumber\\
&=& E(\vl)\pm i B(\vl) - \intlnp [E(\vl')\pm i B(\vl')]
L_P(\bfl,\bflp) - \intlnp \cmb(\bflp) L_{\cmb}(\bfl,\bflp)\,,
\label{E:thetal}
\end{eqnarray}
where
\begin{eqnarray}
\label{E:lfactorforpola} L_P(\bfl,\bflp) &\equiv&
e^{\pm2i(\varphi_{\vl'}-\varphi_\vl)}\len(\bfl-\bflp) \, \left[
(\bfl-\bflp) \cdot \bflp \right] + \frac{1}{2} \intlnpp
e^{\pm2i(\varphi_{\vl'}-\varphi_\vl)} \len(\bflpp)\times\len(\bfl - \bflp
- \bflpp) \, (\bflpp \cdot
\bflp)\left[ (\bflpp + \bflp - \bfl)\cdot \bflp \right] + \ldots \,\nonumber\\
&&\quad -e^{\pm2i(\varphi_{\vl'}-\varphi_\vl)}[\calb(\bfl-\bflp) \pm i
2\rot(\bfl-\bflp)] + \intlnpp
e^{\pm2i(\varphi_{\vl'}-\varphi_\vl)}[\calb(\bflpp) \pm i
2\rot(\bflpp)]\times
[\bflp\cdot(\bfl-\bflp-\bflpp)]\, \len(\bfl-\bflp-\bflpp)\ldots \,  \nonumber\\
&&\quad - e^{\pm
2i(2\varphi_{\vl-\vl'}-\varphi_{\vl'}-\varphi_\vl)}[f_1(\bfl-\bflp) \pm i
f_2(\bfl-\bflp)] + \intlnpp e^{\pm
2i(2\varphi_{\vl''}-\varphi_{\vl'}-\varphi_\vl)}[f_1(\bflpp) \pm i
f_2(\bflpp)]\times [\bflp\cdot(\bfl-\bflp-\bflpp)]\,
\len(\bfl-\bflp-\bflpp)\ldots \,  \nonumber \\
\end{eqnarray}
and
\begin{eqnarray}
L_{\cmb}(\bfl,\bflp) &\equiv&
-e^{\pm2i(\varphi_{\vl-\vl'}-\varphi_{\vl})}[\gamma_1(\bfl-\bflp) \pm i
\gamma_2(\bfl-\bflp)] + \intlnpp
e^{\pm2i(\varphi_{\vl-\vl'}-\varphi_{\vl})}[\gamma_1(\bflpp) \pm i
\gamma_2(\bflpp)]\times [\bflp\cdot(\bfl-\bflp-\bflpp)]\,
\len(\bfl-\bflp-\bflpp)\ldots \,  \nonumber\\
\end{eqnarray}

Similarly for the local coupling systematics, the Taylor expansion of two Stokes polarization parameters including the systematic parameters can be written as
\begin{eqnarray}
\label{eqn:localmodel} [\tQ \pm i \tU]^{obs}(\bn) & = & (1 +
\delta)[Q \pm i U](\bn + \nabla\len(\bn);\sigma) = [Q \pm i U](\bn +
\nabla\len(\bn);\sigma)+ \sigma {\bf p}(\bn) \cdot \nabla [Q \pm i
U](\bn + \nabla\len(\bn);\sigma)\nonumber\\
&& + \sigma [d_1 \pm i d_2](\bn) [\partial_1 \pm
i\partial_2]\cmb(\bn + \nabla\len(\bn);\sigma) + \sigma^2 q(\bn)
[\partial_1 \pm i
\partial_2]^2 \cmb(\bn + \nabla\len(\bn);\sigma)\,\nonumber\\
&=& [Q \pm i U](\bn;\sigma) + \nabla_a \len(\bn) \nabla^a [Q \pm i
U](\bn;\sigma) + \frac{1}{2}\nabla_a \len(\bn) \nabla_b \len(\bn)
\nabla^{a}\nabla^{b}[Q \pm i U](\bn;\sigma)]+ \ldots \nonumber\\
&& + \sigma {\bf p}(\bn) \cdot \nabla [\,[Q \pm i U](\bn;\sigma) +
\nabla_a \len(\bn;\sigma) \nabla^a [Q \pm i U](\bn;\sigma)  \nonumber\\
&& + \frac{1}{2}\nabla_a \len(\bn;\sigma)
\nabla_b \len(\bn;\sigma) \nabla^{a}\nabla^{b}[Q \pm i U](\bn;\sigma)+ \ldots]\nonumber\\
&& + \sigma [d_1 \pm i d_2](\bn) [\partial_1 \pm
i\partial_2][\cmb(\bn;\sigma) + \nabla_a \len(\bn) \nabla^a
\cmb(\bn;\sigma)\nonumber\\
&& + \frac{1}{2}\nabla_a \len(\bn) \nabla_b \len(\bn)
\nabla^{a}\nabla^{b}\cmb(\bn;\sigma)+ \ldots]\,.\nonumber\\
&&+ \sigma ^2 q(\bn) [\partial_1 \pm i\partial_2]^2
[\cmb(\bn;\sigma) + \nabla_a \len(\bn) \nabla^a
\cmb(\bn;\sigma)\nonumber\\
&& + \frac{1}{2}\nabla_a \len(\bn) \nabla_b \len(\bn)
\nabla^{a}\nabla^{b}\cmb(\bn;\sigma)+ \ldots]\,.
\end{eqnarray}

Again performing the harmonic transformation gives
\begin{eqnarray}
[\tilde E(\vl)\pm i \tilde B(\vl)]^{obs} &=& \int d \bn\, [\tQ \pm i \tU]^{obs}(\bn) e^{\mp2i\varphi_\vl} e^{-i \bfl \cdot \bn} \nonumber\\
&=& E(\vl)\pm i B(\vl) - \intlnp [E(\vl')\pm i B(\vl')]
M_P(\bfl,\bflp) - \intlnp \cmb(\bflp) M_{\cmb}(\bfl,\bflp)\,,
\end{eqnarray}
where
\begin{eqnarray}
\label{E:lfactorforpola} M_P(\bfl,\bflp) &\equiv&
e^{\pm2i(\varphi_{\vl'}-\varphi_\vl)}\len(\bfl-\bflp) \, \left[
(\bfl-\bflp) \cdot \bflp \right] + \frac{1}{2} \intlnpp
e^{\pm2i(\varphi_{\vl'}-\varphi_\vl)} \len(\bflpp)\len(\bfl - \bflp -
\bflpp) \, (\bflpp \cdot
\bflp)\left[ (\bflpp + \bflp - \bfl)\cdot \bflp \right] + \ldots \,\nonumber\\
&&\quad  \mp i \sigma \vl'\cdot[p_a(\vl-\vl')\pm i
p_b(\vl-\vl')]e^{\pm2i(\varphi_{\vl'}-\varphi_\vl)} e^{\pm
i\varphi_{\vl-\vl'}}\nonumber\\
&&\quad \pm i \sigma \intlnpp [p_a(\bflpp) \pm i
p_b(\bflpp)]\cdot(\bfl-\bflpp)\, [\bflp\cdot(\bfl-\bflp-\bflpp)]\,
\len(\bfl-\bflp-\bflpp)e^{\pm2i(\varphi_{\vl'}-\varphi_\vl)}e^{\pm
i\varphi_{\vl''}}\ldots \,,
\end{eqnarray}
and
\begin{eqnarray}
M_{\cmb}(\bfl,\bflp) &\equiv& i \sigma \vl'[d_a(\vl-\vl')\pm
i d_b(\vl-\vl')]e^{\pm i(\varphi_{\vl'}+\varphi_{\vl-\vl'}-2\varphi_\vl)}\nonumber\\
&&\quad - i \sigma \intlnpp [d_a(\bflpp) \pm i d_b(\bflpp)]\cdot
[e^{\pm i\varphi_{\vl-\vl'-\vl''}}(\vl-\vl'-\vl'')+ e^{\pm
i\varphi_{\vl'}}\vl']
[\bflp\cdot(\bfl-\bflp-\bflpp)]\, \len(\bfl-\bflp-\bflpp)e^{\pm i\varphi_{\vl''}}e^{\mp2i\varphi_{\vl}}\ldots \,  \nonumber\\
&&\quad +\sigma^2 q(\vl-\vl')\vl'^2
e^{\pm2i(\varphi_{\vl'}-\varphi_\vl)}\nonumber\\
&&\quad - \sigma^2 \intlnpp q(\vl'')
[e^{\pm2i(\varphi_{\vl-\vl'-\vl''}-\varphi_{\vl})}(\vl-\vl'-\vl'')^2+
e^{\pm2i(\varphi_{\vl'}-\varphi_{\vl})}\vl'^2]\,
[\bflp\cdot(\bfl-\bflp-\bflpp)]\, \len(\bfl-\bflp-\bflpp)\ldots \,.
\label{E:sysfilter}
\end{eqnarray}
Using equations ~(\ref{E:thetal})-(\ref{E:sysfilter}), the
window functions for each systematic fields (defined in
Table~\ref{table:geometricd}) can be obtained without any difficulty.

\section{Systematic Contamination in Other Estimators}
\label{otherestimators} In order to keep the paper short and clean,
we focused on the EB estimator, one among six possible quadratic
estimators of CMB fields. In this appendix, we give the results for
EE, TE, and TB estimators.

\subsection{EE estimator}
For CMBPol like experiment, for multiple $l \lesssim 100$, EE
estimator has comparable noise for the
 $C_l^{dd}$ reconstruction (especially in case of noise dominated B-mode detection).
  If one has good signal to noise ratio on E-mode detection, EE estimator can be a better choice compared
  to EB estimator. Here we calculate the systematic contamination for EE estimator. Again,
  we first calculate the connected trispectrum, which can be written in a compact form using the window functions defined in Table~\ref{table:geometricd},
\begin{eqnarray}
\left< \tilde E(\bfl_1)^\obs \tilde E(\bfl_2)^\obs\tilde
E(\bfl_1')^\obs\tilde E(\bfl_2')^\obs\right>_c &=& (2\pi)^2
\delta_\dirac(\vecl_1+\vecl_2+\vecl_1'+\vecl_2')\times \nonumber \\
&& \quad \Bigg\{C_{l_1}^{EE}C_{l_1'}^{EE} \Big<
\len(\vecl_1+\vecl_2)\len(\vecl_1'+\vecl_2') \Big>
W_E(\vl_2,-\vl_1)W_E(\vl_2',-\vl_1') \nonumber \\
&& \quad + \sum^{P-distortion}_{SS'} C_{l_1}^{EE} C_{l_1'}^{EE}
\Big< S(\vecla+\veclb) {S'}(\vecl_1'+\vecl_2')\Big>
W^S_{E}(\vl_2,-\vl_1)W^{S'}_{E}(\vl_2',-\vl_1') \nonumber \\
&& \quad + \sum^{T-leakage}_{SS'} C_{l_1}^{TE} C_{l_1'}^{TE} \Big<
S(\vecla+\veclb) {S'}(\vecl_1'+\vecl_2')\Big>
W^{S}_{E}(\vl_2,-\vl_1)W^{S'}_{E}(\vl_2',-\vl_1') \nonumber \\
&& \quad + C_{l_1}^{EE}C_{l_1'}^{EE} \Big< \len(\vecla+\vecl_2')
\len(\veclb+\vecl_1') \Big> \, W_E(\vl_2,-\vl_1')W_E(\vl_2',-\vl_1) \nonumber \\
&& \quad + \sum^{P-distortion}_{SS'}  C_{l_1}^{EE} C_{l_1'}^{EE}
\Big< S(\vecla+\vecl_2') {S'}(\veclb+\vecl_1')
\Big>\,W^{S}_{E}(\vl_2,-\vl_1')W^{S'}_{E}(\vl_2',-\vl_1) \nonumber\\
&& \quad + \sum^{T-leakage}_{SS'}  C_{l_1}^{TE} C_{l_1'}^{TE} \Big<
S(\vecla+\vecl_2') {S'}(\veclb+\vecl_1')
\Big>\,W^{S}_{E}(\vl_2,-\vl_1')W^{S'}_{E}(\vl_2',-\vl_1) +
Perm.\Bigg\} \,
\end{eqnarray}
Further averaging over the lensing and the systematic field gives
\begin{eqnarray}
\left< \tilde E(\bfl_1)^\obs \tilde E(\bfl_2)^\obs\tilde
E(\bfl_1')^\obs\tilde E(\bfl_2')^\obs\right>_c
&=& (2\pi)^2 \delta_\dirac(\vecl_1+\vecl_2+\vecl_1'+\vecl_2')\times \nonumber \\
&&\quad \Bigg\{ C_{l_1}^{EE}C_{l_1'}^{EE}
\Bigg[C^\pp_{|\vecl_1+\vecl_2|}W_E(\vl_2,-\vl_1)W_E(\vl_2',-\vl_1')
 + C^\pp_{|\vecl_1+\vecl_2'|} W_E(\vl_2,-\vl_1')W_E(\vl_2',-\vl_1)\nonumber \\
&& \quad
+\,\sum^{P-distortion}_{S}C_{|{\vecl_1}+{\vecl_2}|}^{SS}W^{S}_{E}(\vl_2,-\vl_1)W^{S}_{E}(\vl_2',-\vl_1')\,+
\,\sum^{P-distortion}_{S}C_{|{\vecl_1}+{\vecl_2'}|}^{SS}W^{S}_{E}(\vl_2,-\vl_1')W^{S}_{E}(\vl_2',-\vl_1)\Bigg] \nonumber \\
&&\quad + \,C_{l_1}^{TE}C_{l_1'}^{TE}
\Bigg[\,\sum^{T-leakage}_{S}C_{|{\vecl_1}+{\vecl_2}|}^{SS}W^{S}_{E}(\vl_2,-\vl_1)W^{S}_{E}(\vl_2',-\vl_1')\,+ \nonumber \\
&&\quad\,\sum^{T-leakage}_{S}C_{|{\vecl_1}+{\vecl_2'}|}^{SS}W^{S}_{E}(\vl_2,-\vl_1')W^{S}_{E}(\vl_2',-\vl_1)\Bigg]\Bigg\}
+ Perm
 \, , \label{eqn:triEE} \end{eqnarray}
  where we define the lensing E-mode window
function as $W_E(\vl,\vl')\equiv -\vl'\cdot(\vl-\vl')\cos
2(\varphi_{\vl'}-\varphi_{\vl})$. The systematic window functions
$W^{S}_{E}(\vl_1,-\vl_1')$
 for any of those 11 systematics parameters are defined in Table~\ref{table:geometricd}. Hence we obtain the CMB E-mode polarization trispectrum due to
gravitational lensing in the presence of systematic contamination. We define

\begin{eqnarray}
f^{S}_{P}(\bfl_1, \bfl_2) = C_{l_1}^{EE}W^{S}_{E}(\vl_1,\vl_2) +
C_{l_2}^{EE} W^{S}_{E}(\vl_2,\vl_1),
\end{eqnarray}
and
\begin{eqnarray}
f^{S}_{T}(\bfl_1, \bfl_2) = C_{l_1}^{TE}W^{S}_{E}(\vl_1,\vl_2) +
C_{l_2}^{TE} W^{S}_{E}(\vl_2,\vl_1)\,.
\end{eqnarray}
Thus Eq.~(\ref{eqn:triEE}) could be written as
\begin{eqnarray} \label{E:trispectrum}
&&\left< \tilde E(\bfl_1)^\obs \tilde E(\bfl_2)^\obs\tilde
E(\bfl_1')^\obs\tilde E(\bfl_2')^\obs\right>_c =\Bigg[
C_{|\vecla+\veclb|}^\pp f_{EE}(\vecla,\veclb)
f_{EE}(\vecla',\veclb') + C_{|\vecla+\vecla'|}^\pp
f_{EE}(\vecla,\vecla')f_{EE}(\veclb,\veclb') +
C_{|\vecla+\veclb'|}^\pp f_{EE}(\vecla,\veclb')
f_{EE}(\veclb,\vecla')\Bigg] \nonumber \\&&
\quad \quad \quad \quad  \quad \quad \quad \quad
+ \sum^{P-distortion}_{SS'} \Bigg[ C_{|\vecla+\veclb|}^{SS}
f^{S}_{EE}(\bfl, \bflp)f^{S}_{EE}(\bfl, \bflp)+
C_{|\vecla+\vecla'|}^{SS}
f^{S}_{EE}(\vecla,\vecla')f^{S}_{EE}(\veclb,\veclb')  
+ C_{|\vecla+\veclb'|}^{SS} f^{S}_{EE}(\vecla,\veclb')
f^{S}_{EE}(\veclb,\vecla')\Bigg] \nonumber \\&& \quad \quad \quad
\quad \quad \quad  \quad \quad + \sum^{T-leakage}_{SS'} \Bigg[
C_{|\vecla+\veclb|}^{SS} f^{S}_{EE}(\bfl, \bflp)f^{S}_{EE}(\bfl,
\bflp)+ C_{|\vecla+\vecla'|}^{SS}
f^{S}_{EE}(\vecla,\vecla')f^{S}_{EE}(\veclb,\veclb') +
C_{|\vecla+\veclb'|}^{SS} f^{S}_{EE}(\vecla,\veclb')
f^{S}_{EE}(\veclb,\vecla')\Bigg]\,,
\end{eqnarray}
where the permutations now contain 5 additional terms with the
replacement of $(\veclc,\vecld)$ pair by other combination of pairs
$(\vecla,\veclb)$, $(\vecla,\veclc)$, $(\vecla,\vecld)$,
$(\veclb,\veclc)$, $(\veclb,\vecld)$. One can simply Plug in the trispectrum given above to the
deflection angle power spectrum estimation (the final result can be written as a sum of deflection angle power spectrum and noise terms)
\begin{eqnarray} \label{E:1st}
\Big\langle\Big\langle \langle \estEE(\bfL) \cdot \estEE(\bfLp)
\rangle_{\rm{CMB}}  \Big\rangle_{\rm{LSS}}\Big\rangle_{\rm{SYS}} &=&
\frac{\normEE(L)}{L^2} \frac{\normEE(L')}{L'^2} \intl{1} \intl{1'}
\filtEE(\vecla, \veclb) \filtEE(\vecla', \veclb') \nonumber
\\&&\quad \times \left< \tilde E(\bfl_1)^\obs \tilde
E(\bfl_2)^\obs\tilde E(\bfl_1')^\obs\tilde E(\bfl_2')^\obs\right>
 \nonumber \\
&=& (2 \pi)^2 \delta_\dirac(\bfL + \bfLp) \Bigg[
C^{dd}(L)+\noiseEE(L) + \noisePEE(L) + \noiseSEE(L) + ...\Bigg] \, ,
\end{eqnarray}
where $\bfL=\vecla + \veclb$, and $C^{dd}(L)$ is the deflection
angle power spectrum. The terms $\noiseEE(L)$, $\noisePEE(L)$ and
$\noiseSEE(L)$ are the Gaussian noise, first order non-Gaussian
noise, and the first order systematics noise for the EE estimator
respectively.
The systematics noise term  $\noiseSEE(L)$ can be
written as
\begin{eqnarray}
\noiseSEE(L)&=&\frac{\normEE^2(L)}{L^2}\intl{1} \intl{1'}
\filtEE(\vecla, \veclb) \filtEE(\vecla', \veclb')\nonumber \\
&&\quad \Bigg\{ C_{l_1}^{EE}C_{l_1'}^{EE} \Bigg [
\sum^{P-distortion}_{S}C_{|{\vecl_1}+{\vecl_2}|}^{SS}W^{S}_{E}(\vl_2,-\vl_1)W^{S}_{E}(\vl_2',-\vl_1')\,+
\,\sum^{P-distortion}_{S}C_{|{\vecl_1}+{\vecl_2'}|}^{SS}W^{S}_{E}(\vl_2,-\vl_1')W^{S}_{E}(\vl_2',-\vl_1)\Bigg ] \nonumber \\
&&\quad + \,C_{l_1}^{TE}C_{l_1'}^{TE} \Bigg
[,\sum^{T-leakage}_{S}C_{|{\vecl_1}+{\vecl_2}|}^{SS}W^{S}_{E}(\vl_2,-\vl_1)W^{S}_{E}(\vl_2',-\vl_1')\,+
\sum^{T-leakage}_{S}C_{|{\vecl_1}+{\vecl_2'}|}^{SS}W^{S}_{E}(\vl_2,-\vl_1')W^{S}_{E}(\vl_2',-\vl_1)\Bigg
]\Bigg\} \, , \label{eq:NsEE}
\end{eqnarray}
where the systematics window functions $W^{S}_{E}(\vl_1,\vl_2)$ are
given in Table~\ref{table:geometricd},

\subsection{TB and TE estimator}

The instrumental systematic contamination calculation for the TB and
TE estimator is similar to the calculation of EB estimator. At
leading order of $C_{l}^\pp$ and $C_{l}^{SS}$, we have

\begin{eqnarray}
\left< \tilde T(\bfl_1)^\obs \tilde X(\bfl_2)^\obs\tilde
T(\bfl_1')^\obs\tilde X(\bfl_2')^\obs\right>_c &=& (2\pi)^2
\delta_\dirac(\vecl_1+\vecl_2+\vecl_1'+\vecl_2')\times \nonumber \\
&& \quad \Bigg\{C_{l_1}^{TE}C_{l_1'}^{TE} \Big<
\len(\vecl_1+\vecl_2)\len(\vecl_1'+\vecl_2') \Big>
W_X(\vl_2,-\vl_1)W_X(\vl_2',-\vl_1') \nonumber \\
&& \quad + \sum^{P-distortion}_{SS'} C_{l_1}^{TE} C_{l_1'}^{TE}
\Big< S(\vecla+\veclb) {S'}(\vecl_1'+\vecl_2')\Big>
W^S_{X}(\vl_2,-\vl_1)W^{S'}_{X}(\vl_2',-\vl_1') \nonumber \\
&& \quad + \sum^{T-leakage}_{SS'} C_{l_1}^{TT} C_{l_1'}^{TT} \Big<
S(\vecla+\veclb) {S'}(\vecl_1'+\vecl_2')\Big>
W^{S}_{X}(\vl_2,-\vl_1)W^{S'}_{X}(\vl_2',-\vl_1') \nonumber \\
&& \quad + C_{l_1}^{TT}C_{l_1'}^{TT} \Big< \len(\vecla+\vecl_2')
\len(\veclb+\vecl_1') \Big> \, W_X(\vl_2,-\vl_1')W_X(\vl_2',-\vl_1) \nonumber \\
&& \quad + \sum^{P-distortion}_{SS'}  C_{l_1}^{TT} C_{l_1'}^{TT}
\Big< S(\vecla+\vecl_2') {S'}(\veclb+\vecl_1')
\Big>\,W^{S}_{X}(\vl_2,-\vl_1')W^{S'}_{X}(\vl_2',-\vl_1) \nonumber\\
&& \quad + \sum^{T-leakage}_{SS'}  C_{l_1}^{TE} C_{l_1'}^{TE} \Big<
S(\vecla+\vecl_2') {S'}(\veclb+\vecl_1')
\Big>\,W^{S}_{X}(\vl_2,-\vl_1')W^{S'}_{X}(\vl_2',-\vl_1)\Bigg\} \nonumber\\
&=& (2\pi)^2 \delta_\dirac(\vecl_1+\vecl_2+\vecl_1'+\vecl_2')\times \nonumber \\
&&\quad \Bigg\{ C_{l_1}^{TE}C_{l_1'}^{TE}
\Bigg[C^\pp_{|\vecl_1+\vecl_2|}W_X(\vl_2,-\vl_1)W_X(\vl_2',-\vl_1')
 + C^\pp_{|\vecl_1+\vecl_2'|} W_X(\vl_2,-\vl_1')W_X(\vl_2',-\vl_1)\nonumber \\
&& \quad
+\,\sum^{P-distortion}_{S}C_{|{\vecl_1}+{\vecl_2}|}^{SS}W^{S}_{X}(\vl_2,-\vl_1)W^{S}_{X}(\vl_2',-\vl_1')\,+
\,\sum^{P-distortion}_{S}C_{|{\vecl_1}+{\vecl_2'}|}^{SS}W^{S}_{X}(\vl_2,-\vl_1')W^{S}_{X}(\vl_2',-\vl_1)\Bigg] \nonumber \\
&&\quad + \,C_{l_1}^{TT}C_{l_1'}^{TT}
\Bigg[\,\sum^{T-leakage}_{S}C_{|{\vecl_1}+{\vecl_2}|}^{SS}W^{S}_{X}(\vl_2,-\vl_1)W^{S}_{X}(\vl_2',-\vl_1')\,+ \nonumber \\
&&\quad\,\sum^{T-leakage}_{S}C_{|{\vecl_1}+{\vecl_2'}|}^{SS}W^{S}_{X}(\vl_2,-\vl_1')W^{S}_{X}(\vl_2',-\vl_1)\Bigg]\Bigg\}
 \, , \label{eqn:trilens2} \end{eqnarray}
where $X = E$ for the $TE$ estimator, and $X = B$ for the $TB$ estimator. We define the lensing window
function
 $W_B(\vl,\vl')\equiv \vl'\cdot(\vl-\vl')\sin
2(\varphi_{\vl}-\varphi_{\vl'})$ and  $W_E(\vl,\vl')\equiv
-\vl'\cdot(\vl-\vl')\cos 2(\varphi_{\vl}-\varphi_{\vl'})$. Window
$W^{S}_{E}(\vl_1,-\vl_1')$ is the systematics window function for
any of those 11 systematics parameters we defined in
Table~\ref{table:geometricd}. One can simply Plug in the trispectrum given above to the
deflection angle power spectrum estimation (the final result can be written as a sum of deflection angle power spectrum and noise terms),
\begin{eqnarray}
\Big\langle\Big\langle \langle \estTX(\bfL) \cdot \estTX(\bfLp)
\rangle_{\rm{CMB}} } \Big\rangle_{\rm{LSS}}\Big\rangle_{\rm{SYS} &=&
\frac{\normTX(L)}{L^2} \frac{\normTX(L')}{L'^2} \intl{1} \intl{1'}
\filtTX(\vecla, \veclb) \filtTX(\vecla', \veclb') \nonumber \\
&&\quad \times \left< \tilde T(\bfl_1)^\obs \tilde
X(\bfl_2)^\obs\tilde T(\bfl_1')^\obs\tilde X(\bfl_2')^\obs\right>
 \nonumber \\
&=& (2 \pi)^2 \delta_\dirac(\bfL + \bfLp) \Bigg[
C^{dd}(L)+\noiseTX(L) + \noisePTX(L) + \noiseSTX(L) + ...\Bigg] \, ,
\end{eqnarray}
where $\bfL=\vecla + \veclb$, and $C^{dd}(L)$ is the deflection
angle power spectrum. The terms $\noiseTX(L)$, $\noisePTX(L)$ and
$\noiseSTX(L)$ are the Gaussian noise, first order non-Gaussian
noise, and the first order systematics noise for the TB/TE estimator
respectively. The systematics noise term $\noiseSTX(L)$ can be
written as
\begin{eqnarray}
\noiseSTX(L)&=&\frac{\normTX^2(L)}{L^2}\intl{1} \intl{1'}
\filtTX(\vecla, \veclb) \filtTX(\vecla', \veclb')\nonumber \\
&&\quad \Bigg\{ C_{l_1}^{TX}C_{l_1'}^{TX} \Bigg [
\sum^{P-distortion}_{S}C_{|{\vecl_1}+{\vecl_2}|}^{SS}W^{S}_{X}(\vl_2,-\vl_1)W^{S}_{X}(\vl_2',-\vl_1')\,+
\,\sum^{P-distortion}_{S}C_{|{\vecl_1}+{\vecl_2'}|}^{SS}W^{S}_{X}(\vl_2,-\vl_1')W^{S}_{X}(\vl_2',-\vl_1)\Bigg ] \nonumber \\
&&\quad + \,C_{l_1}^{TT}C_{l_1'}^{TT} \Bigg
[\,\sum^{T-leakage}_{S}C_{|{\vecl_1}+{\vecl_2}|}^{SS}W^{S}_{X}(\vl_2,-\vl_1)W^{S}_{X}(\vl_2',-\vl_1')\,+
\sum^{T-leakage}_{S}C_{|{\vecl_1}+{\vecl_2'}|}^{SS}W^{S}_{X}(\vl_2,-\vl_1')W^{S}_{X}(\vl_2',-\vl_1)\Bigg
]\Bigg\}\, , \label{eq:NsTX}
\end{eqnarray}
where the systematics window functions $W^{S}_{X}(\vl_1,\vl_2)$ are
given in Table~\ref{table:geometricd}.


\begin{thebibliography}{99}


\bibitem{Kovac} J. M. Kovac,  et al., Nature 420, 772 (2002).


\bibitem{WMAP} D. N. Spergel, et al., Astrophys. J. Suppl. Ser. 170, 377
(2007).

\bibitem{ZalSel97} U. Seljak, and M. Zaldarriaga, Phys. Rev. Lett. 78, 2054 (1997); M. Zaldarriaga, and U. Seljak, Phys. Rev. D 55, 1830 (1997).

\bibitem{KamKosSte97}
M. Kamionkowski, A. Kosowsky, A. Stebbins,  Phys. Rev. D 55, 7368
(1997); M. Kamionkowski, A. Kosowsky, A. Stebbins, Phys. Rev. Lett.
78, 2058 (1997).

\bibitem{Sel96} U. Seljak, Astrophys. J. 463, 1 (1996).

\bibitem{ZalSel98}
M. Zaldarriaga, and U. Seljak, Phys. Rev. D 58, 023003 (1998); U.
Seljak, and M. Zaldarriaga, Phys. Rev. Lett. 82, 2636 (1999);
 U. Seljak, and M. Zaldarriaga, Phys. Rev. D 60, 043504 (1999);

\bibitem{Hu00b} W. Hu, Phys. Rev. D 62, 043007 (2000).

\bibitem{others}
K. Benabed, F. Bernardeau, and L. Waerbeke, Phys. Rev. D 63, 043501
(2001).

\bibitem{HHS}W. Hu, D. Huterer, and K. M. Smith,  Astrophys. J. 650,
L13 (2006).

\bibitem{smith}K. M. Smith, W. Hu, and M. Kaplinghat, Phys. Rev. D 74, 123002
(2006).

\bibitem{smith2} S. Smith, A. Challinor, and G. Rocha, Phys. Rev. D 73, 023517 (2006).
(2006).

\bibitem{viviana} V. Acquaviva, and C. Baccigalupi, Phys. Rev. D 74, 103510
(2006).

\bibitem{lensinginf} E. Calabrese, et. al, Phys. Rev. D 77, 123531
(2008).

\bibitem{secondaryBmode} D. Sarkar, P. Serra, A. Cooray, K. Ichiki,
and D. Baumann, Phys. Rev. D 77, 103515 (2008).

\bibitem{HuOkamoto} W. Hu, Astrophys. J., 557, L79 (2001); W. Hu, and T. Okamoto, Astrophys. J. 574, 566
(2002); T. Okamoto, and W. Hu, Phys. Rev. D  67, 083002 (2003).

\bibitem {ZalSel99} M. Zaldarriaga, and U. Seljak, Phys. Rev. D 59,
123507 (1999); M. Zaldarriaga, Phys. Rev. D 62, 063510 (2000).

\bibitem{GuzSelZal00} J. Guzik, U. Seljak, and M. Zaldarriaga, Phys. Rev. D 62, 043517 (2000).

\bibitem{Hirata} C. M. Hirata, and U. Seljak, Phys. Rev. D 67,
043001 (2003); C. M. Hirata, and U. Seljak, Phys. Rev. D 68, 083002
(2003).

\bibitem{Hirata2} U. Seljak, and C. M. Hirata, Phys. Rev. D 69, 043005 (2004).

\bibitem{Kesden2} M. Kesden, A. Cooray, and M. Kamionkowski, Phys. Rev. Lett. 89, 011304 (2002).

\bibitem{Kesden} M. Kesden, A. Cooray, and M. Kamionkowski, Phys. Rev. D 67,
123507 (2003).

\bibitem{Cooray} A. Cooray, Phys. Rev. D 65, 063512 (2002); A. Cooray, and M. Kesden, New Astron. 8, 231 (2003); A. Cooray, Phys. Rev. D 66, 103509 (2002).

\bibitem{Hu01c} W. Hu, Phys. Rev. D 65, 023003 (2002).

\bibitem{CMBpol} http://cmbpol.uchicago.edu/; D. Baumann, et al.,
arXiv:0811.3911; K. M. Smith, et al., arXiv:0811.3916;  M.
Zaldarriaga, et al., arXiv:0811.3918; D. Baumann, et al.,
arXiv:0811.3919; J. Dunkley, et al., arXiv:0811.3915; A. A. Fraisse,
et al., arXiv:0811.3920.


\bibitem{EBEX} P. Oxley et al., Proc. SPIEInt. Soc. Opt. Eng., 5543, 320 (2004), arXiv:astro-ph/0501111;
http://groups.physics.umn.edu/cosmology/ebex/.

\bibitem{Spider} C. J. MacTavish, et al., arXiv:0710.0375; B. P. Crill, et al. arXiv
0807.1548.

\bibitem{QUIJOTE} J.A. Rubi\~no-Mart\'in, et al., arXiv:0810.3141.

\bibitem{Task} J. Bock, et al., arXiv:astro-ph/0604101.

\bibitem{EPIC} J. Bock, et al., arXiv:0805.4207.

\bibitem{Capmap} C. Bischoff, et al. Astrophys. J. 684, 771 (2008).


\bibitem{experiments} L. Verde, H. Peiris, and R. Jimenez, JCAP, 0601,
019 (2006); M. Kaplinghat, New Astronomy Reviews, 47, 893 (2003).


\bibitem{HHZ}W. Hu, M.~M. Hedman, \& M. Zaldarriaga, Phys. Rev. D, 67,
043004 (2003).

\bibitem{DCJ} D. O'Dea, A. Challinor, \& B.~R. Johnson, MNRAS, 376,
1767 (2007).

\bibitem{Shimon} M. Shimon, B. Keating, N. Ponthieu, and E. Hivon, Phys. Rev. D 77,
083003 (2008).

\bibitem{Shimon2} N. J. Miller, M. Shimon, B. G. Keating,
arXiv:0806.3096.

\bibitem{Bunn} E. F. Bunn, Phys. Rev. D 75, 083517 (2007).

\bibitem{Beammismatch} C. Rosset, et al. A\&A, 464, 405 (2006).

\bibitem{modulation} M. L. Brown, et al., arXiv:0809.4032.

\bibitem{beamasym} M. A. J. Ashdown, et al., arXiv:0806.3167.


\bibitem{CMBFAST} U. Seljak, and M. Zaldarriaga,  Astrophys. J. 469, 437
(1996).

\bibitem {review} A. Lewis, and A. Challinor, Phys. Rept. 429, 1
(2006).

\bibitem{SZ} R. A. Sunyaev, and Ya. B. Zel'dovich, MNRAS, 190, 413 (1980).

\bibitem{ISW} R. K. Sachs, and A. M. Wolfe, Astrophys. J. 147, 73
(1967).

\bibitem{ksz} M. A. Riquelme, and D. N. Spergel,  Astrophys. J. 661, 672
(2007).

\bibitem{nonGaussianity} J. Lesgourgues, M. Liguori, S. Matarrese,
and A. Riotto, Phys. Rev. D 71, 103514 (2005).


\bibitem{trispectrum} T. Okamoto, and W. Hu, Phys. Rev. D 66, 063008
(2002); W. Hu, Phys. Rev. D 64, 083005 (2001).


\bibitem{fullsky} A. Challinor, and A. Lewis, Phys. Rev. D 71,
103010 (2005).




\end{thebibliography}
\end{document}